\documentclass{mn2e}
\usepackage{natbib2,natbibmnfix}
\usepackage{astrojournals}
\usepackage{epsfig}
\usepackage{amssymb}
\usepackage{multirow}
\usepackage{multicol}
\usepackage{color}
\usepackage[varg]{txfonts}
\usepackage{url}

\def \mathbi#1{\textbf{\em #1}}

\date{}
\pagerange{\pageref{firstpage}--\pageref{lastpage}}
\pubyear{2015}

\begin{document}
\label{firstpage}

\title[Companions tear accretion discs]{Tearing up a misaligned accretion disc with a binary companion}
\author[Do\u{g}an, Nixon, King \& Price]{
Suzan~Do\u{g}an$^{1,2}$\thanks{suzan.dogan@ege.edu.tr}, Chris~Nixon$^{3}$\thanks{chris.nixon@jila.colorado.edu}\thanks{Einstein Fellow}, Andrew~King$^2$ \& Daniel~J.~Price$^4$\vspace{0.1in}\\
$^1$University of Ege, Department of Astronomy \& Space Sciences, Bornova, 35100, ${\dot {\rm I}}$zmir, Turkey\\
$^2$Department of Physics and Astronomy, University of Leicester, University Road, Leicester LE1 7RH, UK\\
$^3$JILA, University of Colorado \& NIST, Boulder CO 80309-0440, USA\\
$^4$Monash Centre for Astrophysics (MoCA), School of Mathematical Sciences, Monash University, Vic. 3800, Australia}
\maketitle

\begin{abstract}
Accretion discs are common in binary systems, and they are often found to be misaligned with respect to the binary orbit. The gravitational torque from a companion induces nodal precession in misaligned disc orbits. We calculate whether this precession is strong enough to overcome the internal disc torques communicating angular momentum. For typical parameters precession wins: the disc breaks into distinct planes that precess effectively independently. We run hydrodynamical simulations to check these results, and confirm that disc breaking is widespread and generally enhances accretion on to the central object. This applies in many cases of astrophysical accretion, e.g. supermassive black hole binaries and X--ray binaries.
\end{abstract}


\section{Introduction}
\label{intro}
Accretion discs (e.g. \citealt{PR1972}; \citealt{Pringle1981}; \citealt{Franketal2002}) appear in many astrophysical systems. In most cases these discs are probably not completely axisymmetric. Discs may be locally tilted by the Lense--Thirring effect of a central misaligned black hole \citep{BP1975}, by radiation (\citealt{Pringle1996,Pringle1997}) or by the gravity of a companion \citep[e.g.][]{LO2000}. Discs in or around supermassive black hole (SMBH) binaries formed by galaxy mergers may be misaligned with respect to the binary orbit through the chaotic nature of AGN accretion (\citealt{KP2006,KP2007}). The effect on the disc in all these cases is similar. The lack of symmetry produces a torque on misaligned rings of gas which makes their orbits precess differentially. Given a sufficiently strong viscosity communicating the precession between the rings, the disc warps. \cite{PP1983} showed that warps can propagate in two distinct regimes: wave--like for $\alpha \lesssim H/R$, and diffusive for $\alpha \gtrsim H/R$ where $\alpha$ is the Shakura \& Sunyaev dimensionless viscosity parameter \citep{SS1973} and $H/R$ is the disc angular semi--thickness. In this paper we focus on diffusive systems with $\alpha > H/R$, which typically holds for accretion discs around black holes.

For diffusive discs subject to differential precession, the expected evolution is that dissipation through viscosity allows the inner parts of the disc to align, joined by a smoothly warped region to the still misaligned outer parts. In the case of a misaligned disc around a spinning black hole, this is often called the Bardeen--Petterson effect. Until recently it was implicitly assumed that this is what always occurs in a diffusive disc, i.e. that the internal disc torques would always be able to communicate the precession. However an analytic study by \cite{Ogilvie1999} pointed out that the effect of a disc warp was to {\it weaken} the communication of angular momentum in the disc, and so weaken the disc's ability to hold itself together. Although his study assumed a locally isotropic viscous process, there appears no reason to assume this behaviour does not hold for a viscosity driven by turbulence, such as the magnetorotational instability \citep[MRI;][]{BH1991}. Indeed for a viscosity driven by magnetic fields it is likely that the vertical viscosity, associated with keeping the disc flat, is weaker than that of an isotropic model, as the vertical gas shear is probably oscillatory whereas the azimuthal shear grows secularly as gas parcels continually move apart \citep{Pringle1992}.

We note that it is conventional to use the term {\it isotropic viscosity}, but that this can be misleading. For a warped disc this term means that the horizontal and vertical shear in the warp are assumed to be damped by viscous dissipation at the same average rates \citep[cf.][eq. 40]{LP2007}. This assumption, made by \cite{PP1983} and \cite{Ogilvie1999}, leads to the result that the azimuthal shear viscosity $\nu_1 \propto \alpha$, but that the vertical viscosity $\nu_2 \propto 1/\alpha$, quite contrary to any naive belief that $\nu_1$ and $\nu_2$ might end up roughly equal. This apparently paradoxical result comes about because a small value of isotropic viscosity $\alpha$ allows a large resonant radial velocity $v_R$ in the warp: the viscous dissipation rate goes as $\alpha v_R^2 \propto 1/\alpha$ and so {\it increases} as $\alpha$ decreases. A warped--disc isotropic viscosity explicitly does not assume that $\nu_1 = \nu_2$, as was the case in the early work on warped discs by e.g. \cite{BP1975}.

An $\alpha$ viscosity that acts isotropically, as described above, appears to hold for viscosity arising from turbulence induced by the MRI \citep{Kingetal2013}. However, very little effort has been directed towards estimating the effective viscosities in magnetised warped discs. \cite{Torkelssonetal2000} performed shearing box calculations to follow the decay of an imposed epicyclic shearing motion, which mimics a warp. Their results are in approximate agreement with an isotropic viscosity (which predicts $\alpha_2 \sim 1/\alpha$; \citealt{PP1983}). Further, \cite{Ogilvie2003} developed an analytic model for the dynamical evolution of magnetorotational turbulent stresses which predicts agreement with the conclusions of \cite{Torkelssonetal2000}. Both of these investigations, numerical {\it and} analytic, allow the effective viscosity from magneto--hydrodynamic (MHD) turbulence to be anisotropic, but both conclude that it is close to isotropic.

The realization that the viscosity may not be strong enough to hold the disc together has significant implications. If the viscosity is too weak, or the external torque on the disc too strong, the disc may instead break into distinct planes with only tenuous gas flows between them \citep{NK2012}. If in addition these planes are sufficiently inclined to the axis of precession,
they can precess until they are
partially counterrotating, promoting angular momentum cancellation and rapid infall -- disc tearing. Tearing occurs in discs inclined to the spin of a central black hole \citep{Nixonetal2012b} and in a circumbinary disc around a misaligned central binary system \citep{Nixonetal2013}.

In this paper we want to find out if tearing can happen {\it inside} a binary, i.e. if a disc around one component can be disrupted by the perturbation from a companion. This would have significant implications for all binary systems: e.g. fuelling SMBH during the SMBH binary phase (cf. Figs. 6 \& 7 of \citealt{Nixonetal2013}) and accretion outbursts in X-ray binaries. In the Lense-Thirring and circumbinary disc cases, disc breaking starts from the inside and works its way outwards. But we also want to know if breaking and tearing can instead start from the {\it outer} edge of a disc internal to a binary, and work its way inwards. To do this we consider binary systems with an initially planar disc around one component, misaligned with respect to the (circular) binary orbit. Following the methods of our earlier papers on disc tearing, we first compare the disc precession torque with the disc viscous torque to determine whether the disc should warp or break. Then we check our findings by comparing this result with hydrodynamical simulations.

\section{Tearing up the disc}
\label{tearing}
\begin{figure}
  \includegraphics[angle=0,width=\columnwidth]{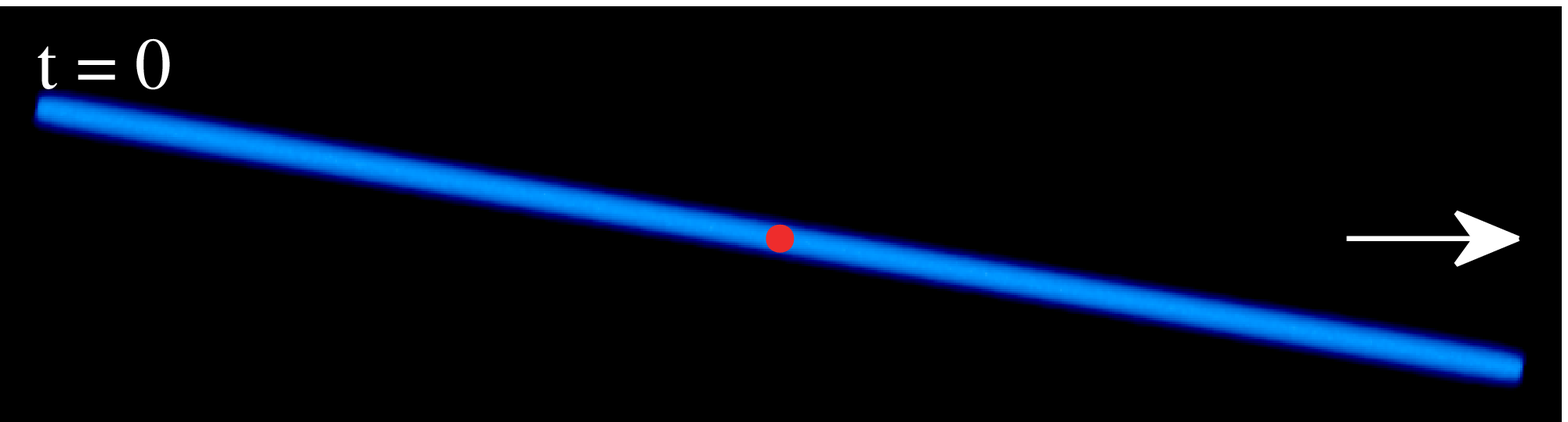}
  \includegraphics[angle=0,width=\columnwidth]{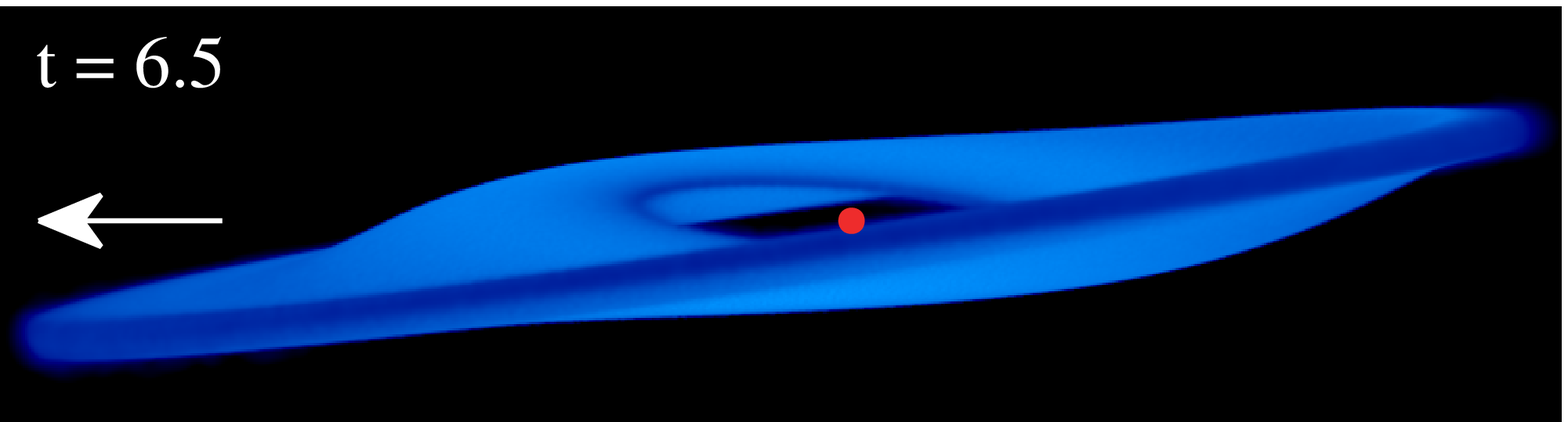}
  \caption{3D surface rendering of the warped disc after 0 and 6.5 binary orbits. The disc was initially inclined at $10^{\circ}$ to the binary plane with no warp. The disc is viewed along the binary orbital plane and the arrow points in the direction of the companion.}
  \label{tilt10}
\end{figure}

\begin{figure*}
  \includegraphics[angle=0,width=0.45\textwidth]{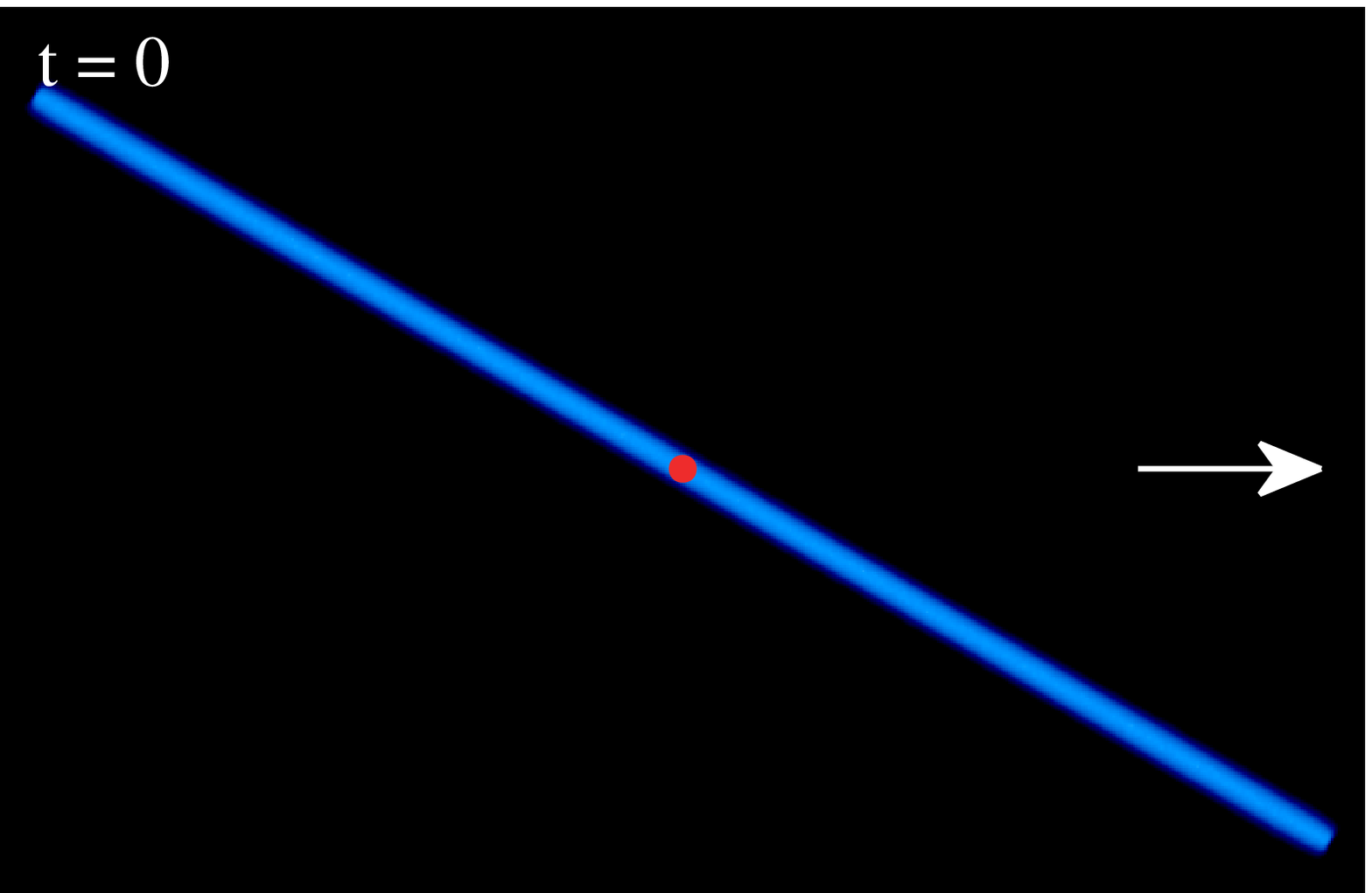}
  \includegraphics[angle=0,width=0.45\textwidth]{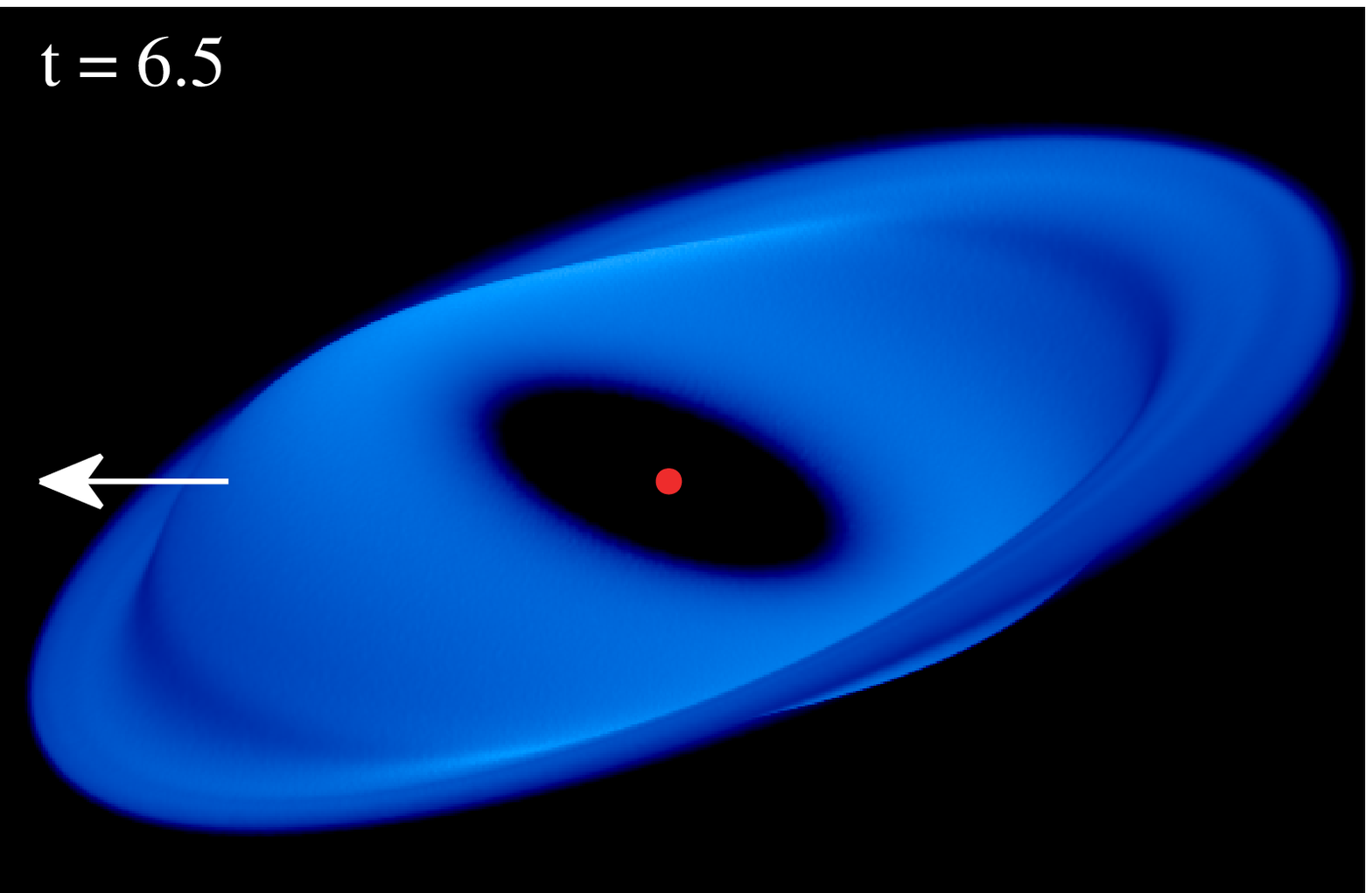}
  \includegraphics[angle=0,width=0.45\textwidth]{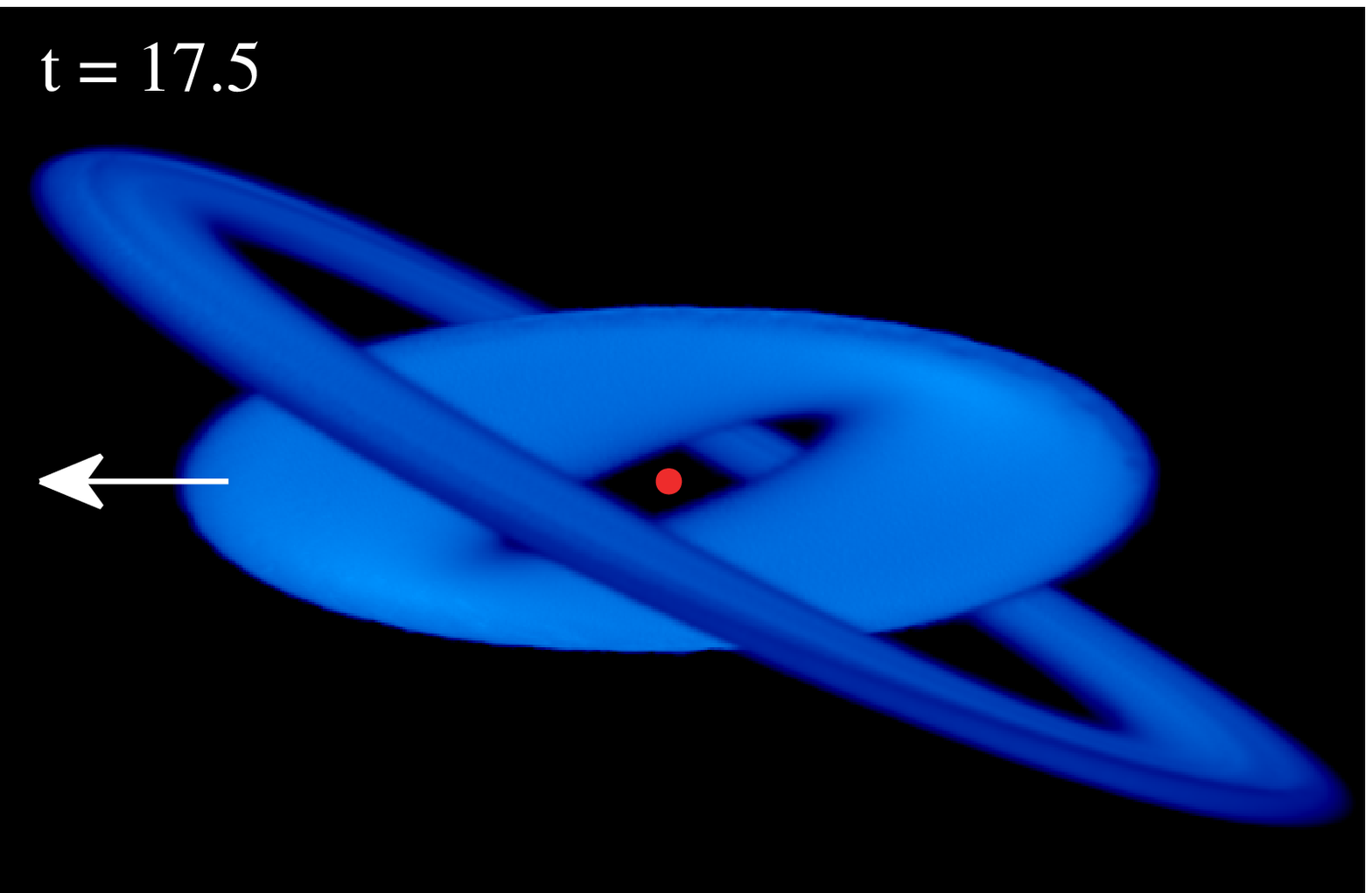}
  \includegraphics[angle=0,width=0.45\textwidth]{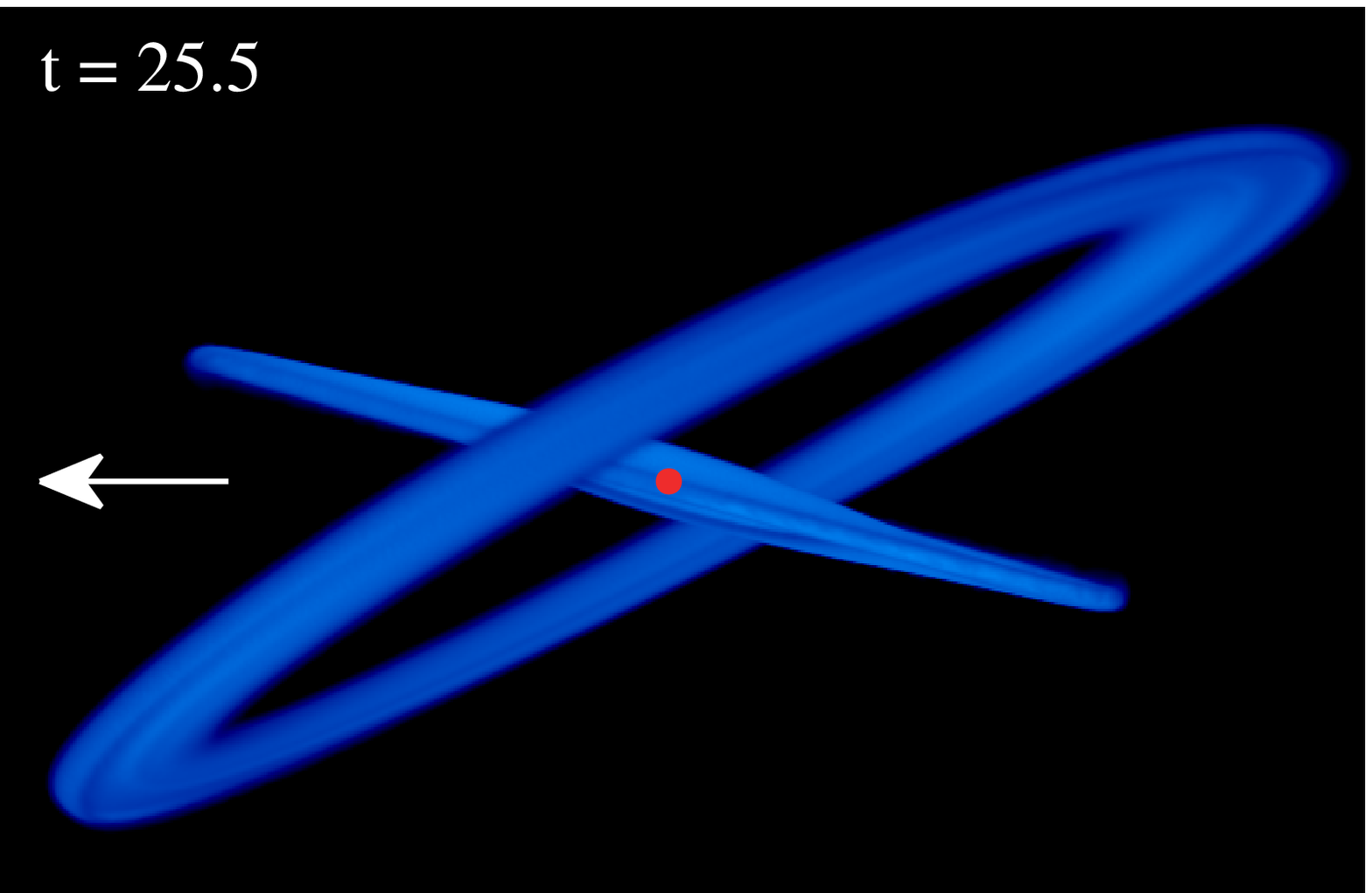}
  \caption{3D surface rendering of the disc which was initially inclined at $30^{\circ}$ to the binary plane with no warp. These snapshots are taken after 0, 6.5, 17.5 and 25.5 binary orbits. The disc is viewed along the binary orbital plane and the arrow points in the direction of the companion. In this simulation the disc breaks into two distinct planes after $\sim$8 binary orbits.}
  \label{tilt30}
\end{figure*}

\begin{figure*}
  \includegraphics[angle=0,width=0.4\textwidth]{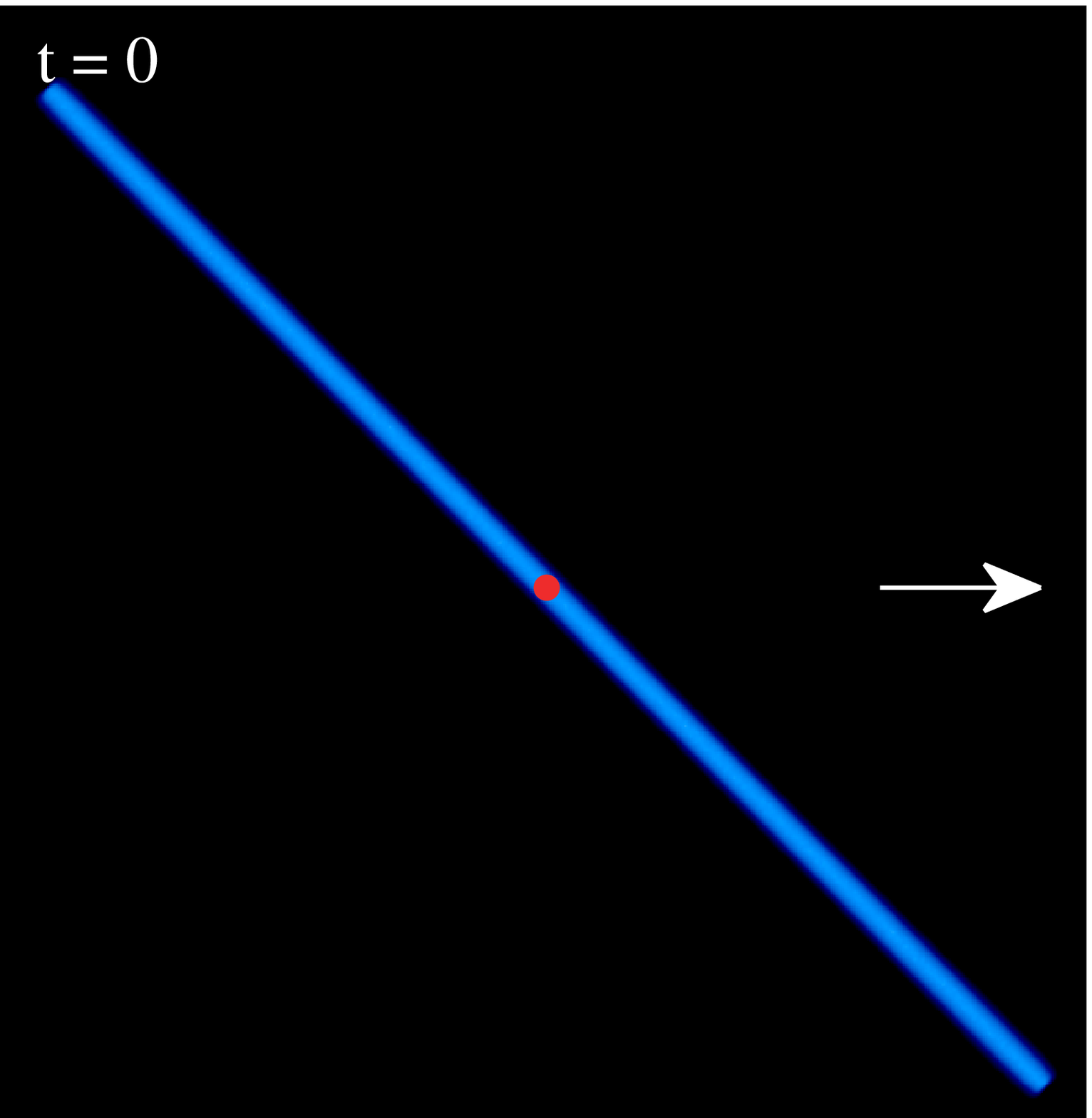}
  \includegraphics[angle=0,width=0.4\textwidth]{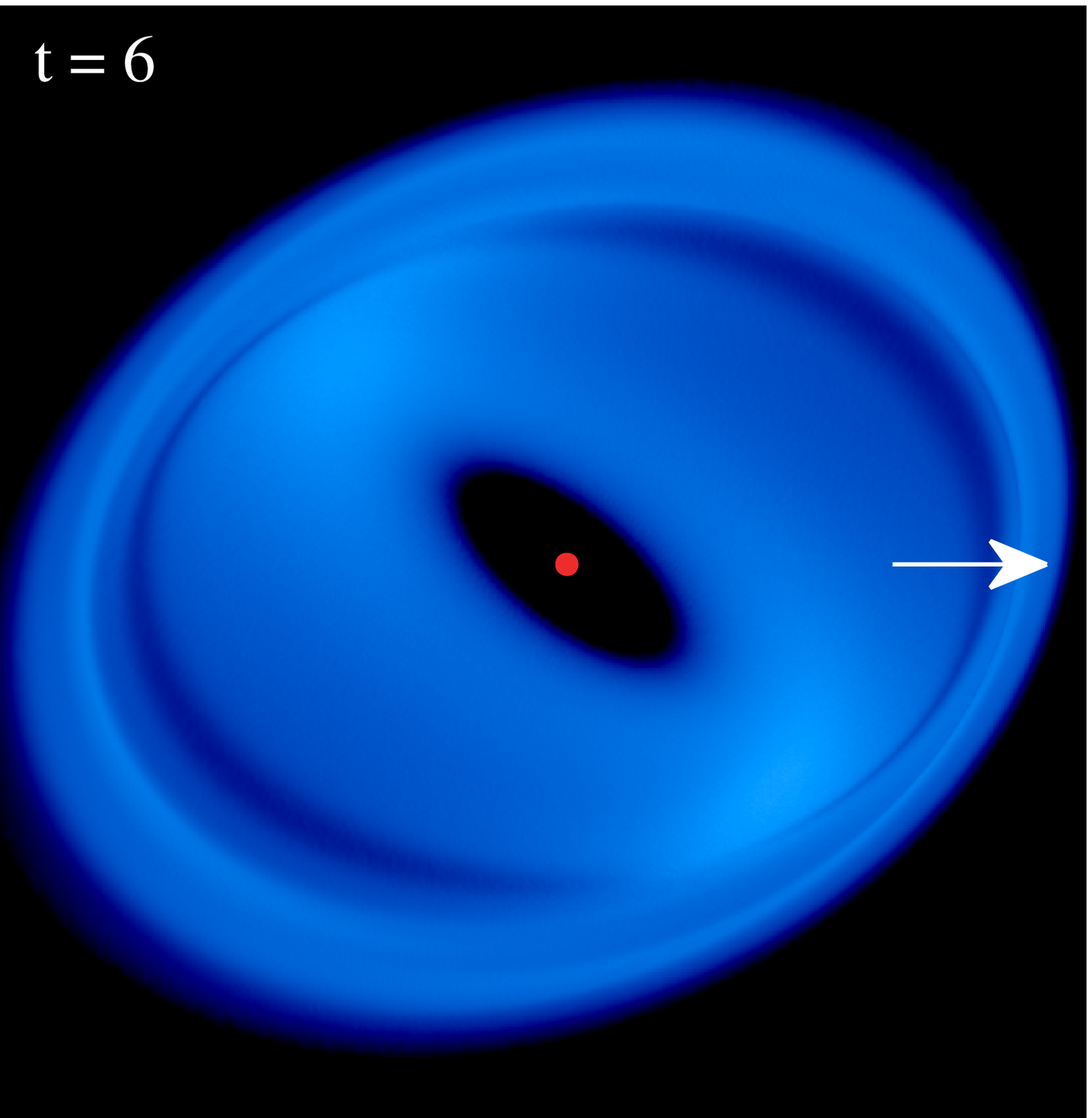}
  \includegraphics[angle=0,width=0.4\textwidth]{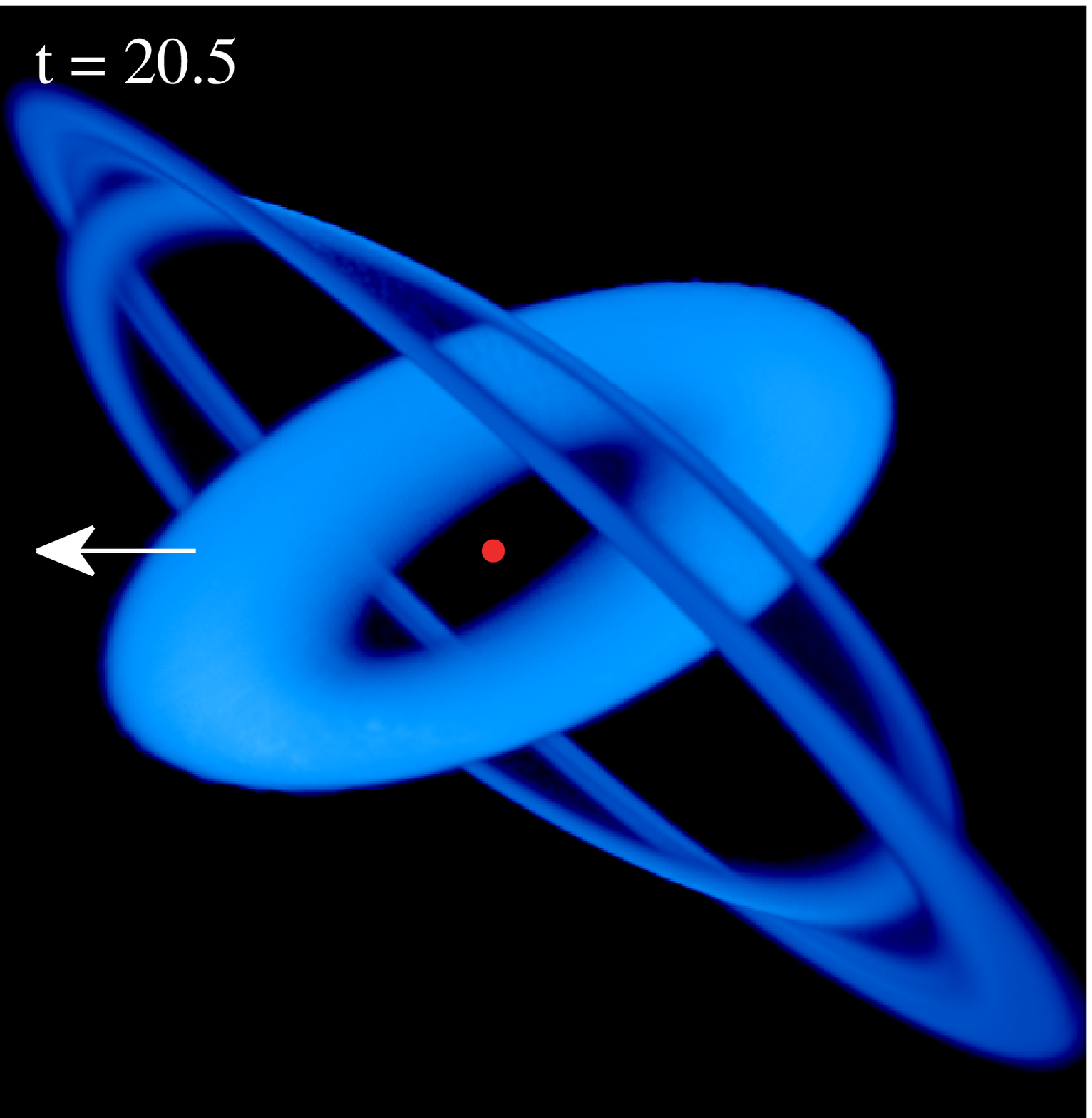}
  \includegraphics[angle=0,width=0.4\textwidth]{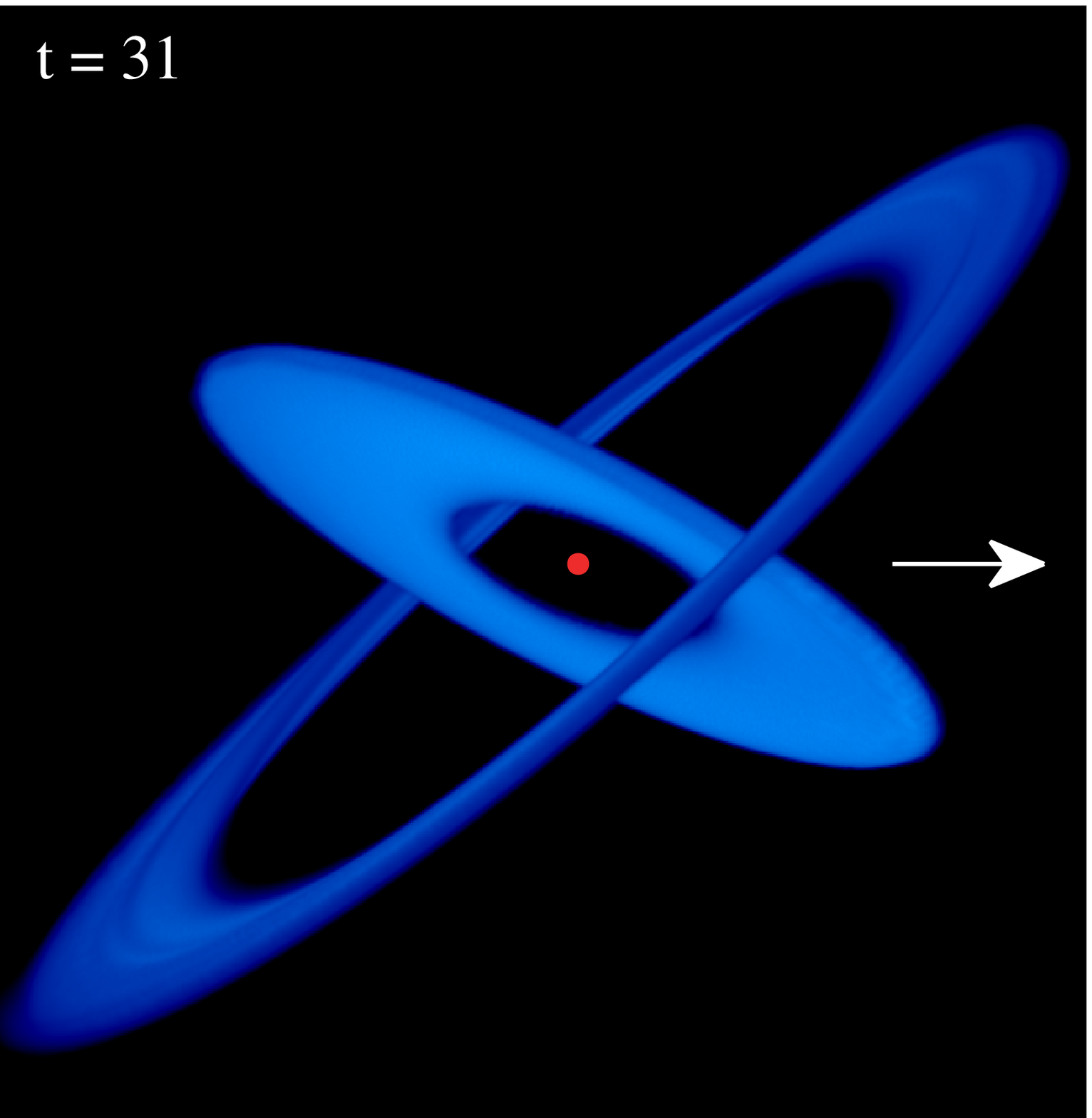}
  \caption{3D surface rendering of the disc which was initially inclined at $45^{\circ}$ to the binary plane with no warp. These snapshots are taken after 0, 6, 20.5 and 31 binary orbits. The disc is viewed along the binary orbital plane and the arrow points in the direction of the companion. In this simulation the disc initially breaks into two distinct planes after $\sim$7 binary orbits.   Then a third ring is broken off, but quickly interacts with the outer ring.}
    \label{tilt45}
\end{figure*}

The disc precession caused by the presence of a binary companion is retrograde, and has frequency \citep{Bateetal2000}
\begin{equation}
\Omega_{\rm p} = \frac{3}{4}\frac{M_{\rm 0}}{M_{\rm 1}}\left(\frac{R}{a}\right)^3\Omega\cos\theta.
\label{omprec}
\end{equation}
Here $\theta$ is the inclination angle between the disc and the binary, $M_0$ \& $M_1$ are the masses of each component of the binary with the disc around $M_1$, $a$ is the binary separation, $R$ (assumed $\ll a$) is the disc radius, and $\Omega = (GM_1/R^3)^{1/2}$ is the disc orbital frequency.

We get an idea of whether the disc tears by estimating the disc precession frequency in a typical case. The disc cannot extend past the Roche lobe radius (more precisely the tidal truncation radius $R_{\rm tide} \sim 0.87 R_{\rm RL}$; \citealt{Franketal2002}), so we take $M_{\rm 1} = M_{\rm 0}$ and $R_{\rm max} \approx 0.35a$. Putting this into (\ref{omprec}) gives
\begin{equation}
\Omega_{\rm p,max} \approx 0.03\Omega\cos\theta.
\end{equation}
So in this case the precession time is only $\sim 30$ dynamical times, suggesting that tearing is possible, as the viscous communication in the disc is likely to be significantly slower than this.

We expect the disc to break when the precession induced in the disc is stronger than any internal communication in the disc. This communication can be due to the usual planar disc viscosity ($\nu_1$), the viscosity arising from vertical shear in a warped disc ($\nu_2$) or pressure. In the simulations presented here we are focusing on the regime with $\alpha > H/R$ and therefore the communication due to pressure is small and we return to this point in Section~\ref{discussion}.

We can write the magnitudes of the viscous torques per unit area as \citep{PP1983}
\begin{equation}
\label{vtorq}
\left|G_{\nu_1}\right| = \frac{3\pi\nu_1\Sigma R^2\Omega}{2\pi R H}
\end{equation}
and
\begin{equation}
\label{gnu2}
\left|G_{\nu_2}\right| = \frac{2\pi R\Sigma R^2\Omega\frac{1}{2}\nu_2\left|\partial \mathbi{l}/\partial R\right|}{2\pi R H}\,.
\end{equation}
Here $\Sigma$ is the disc surface density and $\mathbi{l}$ is the unit angular momentum vector. For Keplerian rotation and a \cite{SS1973} viscosity $\nu_i = \alpha_i H^2 \Omega$ we can write the total as
\begin{equation}
\left|G_{\rm total}\right| = \left|G_{\nu_1}\right| + \left|G_{\nu_2}\right| = \frac{\Sigma R^2 \Omega^2}{2}\frac{H}{R}\left[3\alpha_1 + \alpha_2\left|\psi\right|\right]
\end{equation}
where $|\psi|$ is the warp amplitude and defined as $\left|\psi\right| = R\left|\partial \mathbi{l}/\partial R\right|$ \citep{Ogilvie1999}.

We can compare this to the magnitude of the precession torque per unit area
\begin{equation}
\label{gp}
\left|G_{\rm p}\right| = \left|{\bf \Omega}_{\rm p}\times\mathbi{L} \right| = \frac{3}{4}\frac{M_{\rm 0}}{M_{\rm 1}}\left(\frac{R}{a}\right)^{3}\Sigma R^2
\Omega^2\cos\theta\sin\theta
\end{equation}
to give an idea of where in the disc we expect breaking to occur. Here \textbf{\emph{L}} is the angular momentum density vector. To break the disc the precession must be stronger than its viscous communication, i.e. $\left|G_{\rm p}\right| \gtrsim \left|G_{\rm total}\right|$, giving
\begin{equation}
\label{rbreakt}
R_{\rm break} \gtrsim \left[\frac{4\left(\alpha_1 + \frac{\alpha_2}{3}\left|\psi\right|\right)}{\sin2\theta}\frac{H}{R} \frac{M_{\rm 1}}{M_{\rm 0}}\right]^{1/3} a.
\end{equation}
This break radius accounts for both the azimuthal and vertical viscosities in a warped disc. In contrast, the previous disc tearing papers \citep{Nixonetal2012b,Nixonetal2013} used $\alpha_1=\alpha$ and considered the initial conditions of a flat disc with $\left|\psi\right|=0$.

It is not straightforward to evaluate (\ref{rbreakt}) as both $\alpha_1
$ and $\alpha_2$ are strong functions of the warp amplitude $\left|\psi\right|$ \citep{Ogilvie1999,Ogilvie2000} and the warp amplitude itself is unknown before performing a full 3D calculation of the disc evolution. In previous work it has sufficed to conservatively use $\alpha_1=\alpha$, but to exclude the $\alpha_2$ term. For large $\alpha \gtrsim 0.1$ this is reasonable, but for smaller $\alpha$ the vertical viscosity is clearly important. Proceeding with the method of the earlier papers we get
\begin{equation}
R_{\rm break} \gtrsim \left(\frac{4\alpha}{\sin2\theta}\frac{H}{R} \frac{M_{\rm 1}}{M_{\rm 0}}\right)^{1/3} a,
\label{rbreak}
\end{equation}
but we caution that this equation is not relevant for $\alpha \ll 0.1$ and small inclination angles where the strong vertical viscosity can result in rapid disc alignment. In such cases the total internal torque must be considered (\ref{rbreakt}), but we note that this is not trivial to evaluate beforehand.

We can evaluate (\ref{rbreak}) for typical parameters, giving
\begin{equation}
R_{\rm break} \gtrsim 0.16 a \left(\frac{\alpha}{0.1}\right)^{1/3} \left(\frac{H/R}{0.01}\right)^{1/3} \left(\frac{M_1}{M_0}\right)^{1/3} \left(\sin2\theta\right)^{-1/3}.
\end{equation}
This disc tearing criterion is equivalent to requiring a minimum inclination of the disc to the binary orbit, $\theta_{\rm min}$, defined by
\begin{equation}
\sin2\theta_{\rm min} \gtrsim 4\alpha\frac{H}{R}\frac{M_{\rm 1}}{M_{\rm 0}}\left(\frac{a}{R_{\rm break}}\right)^3.
\label{theta}
\end{equation}
We can simplify this formula in two limits. If the disc is around the less massive component we have $M_{\rm 1} < M_{\rm 0}$ and the tidal limit on the disc size requires
\begin{equation}
{a\over R_{\rm break}} > 2.5\left({M\over M_{\rm 1}}\right)^{1/3},
\label{tide}
\end{equation}
where $M = M_{\rm 1} + M_{\rm 0}$ is the total binary mass, so (\ref{theta}) becomes
\begin{equation}
\sin2\theta \gtrsim 0.06 \left(\frac{\alpha}{0.1}\right)\left(\frac{H/R}{0.01}\right) \ \ \ \ \ \ \ \ \ \ \ \ \ \ \ \ \ \ \ \ \ \ \ \ \ \ \ (M_{\rm 1} < M_{\rm 0})
\label{theta2}
\end{equation}
since $M \simeq M_{\rm 0}$ in this case.

If instead the disc is around the more massive binary component we have $M_{\rm 1} > M_{\rm 0}$ and the disc size is approximately $0.6a$ \citep{AL1994}. In this case, breaking occurs if
\begin{equation}
\sin2\theta \gtrsim 0.18 \left(\frac{\alpha}{0.1}\right)\left(\frac{H/R}{0.01}\right)\left(\frac{M_1/M_0}{10}\right) \ \ \ \ \ \ \ \ \ \  (M_{\rm 1} > M_{\rm 0})
\label{theta3}
\end{equation}

For typical black hole disc parameters $\alpha = 0.1$, $H/R \lesssim 10^{-2}$ almost all discs with a reasonable misalignment should break (cf. \ref{theta2}, \ref{theta3}). However, a very large mass ratio $M_{\rm 1}/M_{\rm 0} \gg 1$ makes the perturbation by the smaller black hole so weak that breaking would occur only after a very long interval.

For X--ray binaries breaking can clearly be avoided in some cases, but probably occurs in others. First, if mass is transferred by Roche lobe overflow, the accretion disc forms closely aligned to the binary plane. So to get any disc inclination to the binary plane\footnote{Note that the binary orbital plane and the spin plane of the black hole may be misaligned and so a disc aligned to the binary plane could still experience Lense--Thirring tearing \citep[e.g.][]{NS2014}.} in a close stellar--mass binary requires a torque to tilt the disc out of the plane. Here Her X-1 is interesting, as this system is known to have a tilted precessing disc \citep{Katz1973}, which sets limits on the viscosity coefficient $\alpha$ \citep[cf.][]{Kingetal2013}. The disc tilt is plausibly attributed to radiation warping (\citealt{Petterson1977a}; \citealt{Petterson1977b}; \citealt{Pringle1996}; \citealt{WP1999}; \citealt{OD2001}) provided that the mass input occurs at small disc radii. \cite{WP1999} estimate $\alpha\simeq 0.3, H/R \simeq 0.04$, and $M_{\rm 1}/M_{\rm 0} \simeq 0.5, R/a \simeq 0.24$. From (\ref{theta}) these give the requirement $\sin2\theta > 1.7$ for breaking to occur, which is of course impossible. This is reassuring, as the disc in Her X-1 appears to precess as a single body. However, slightly larger or thinner discs, or ones with lower viscosity, could easily have values of $\sin2\theta$ allowing for disc breaking.

We note that $R_{\rm break}$ is the radius {\it outside} which we expect the disc to break. This is the opposite of the Lense-Thirring \citep{Nixonetal2012b} and circumbinary \citep{Nixonetal2013} cases, and raises new possibilities. If for example a disc ring broken from the outer edge contained more angular momentum than everything inside, the outer disc might be able to sweep the entire inner disc in to the accretor and leave behind a single misaligned ring. This would presumably spread viscously and possibly repeat the process.

\section{Simulations}
\label{sims}

\begin{figure*}
  \includegraphics[angle=0,width=0.35\textwidth]{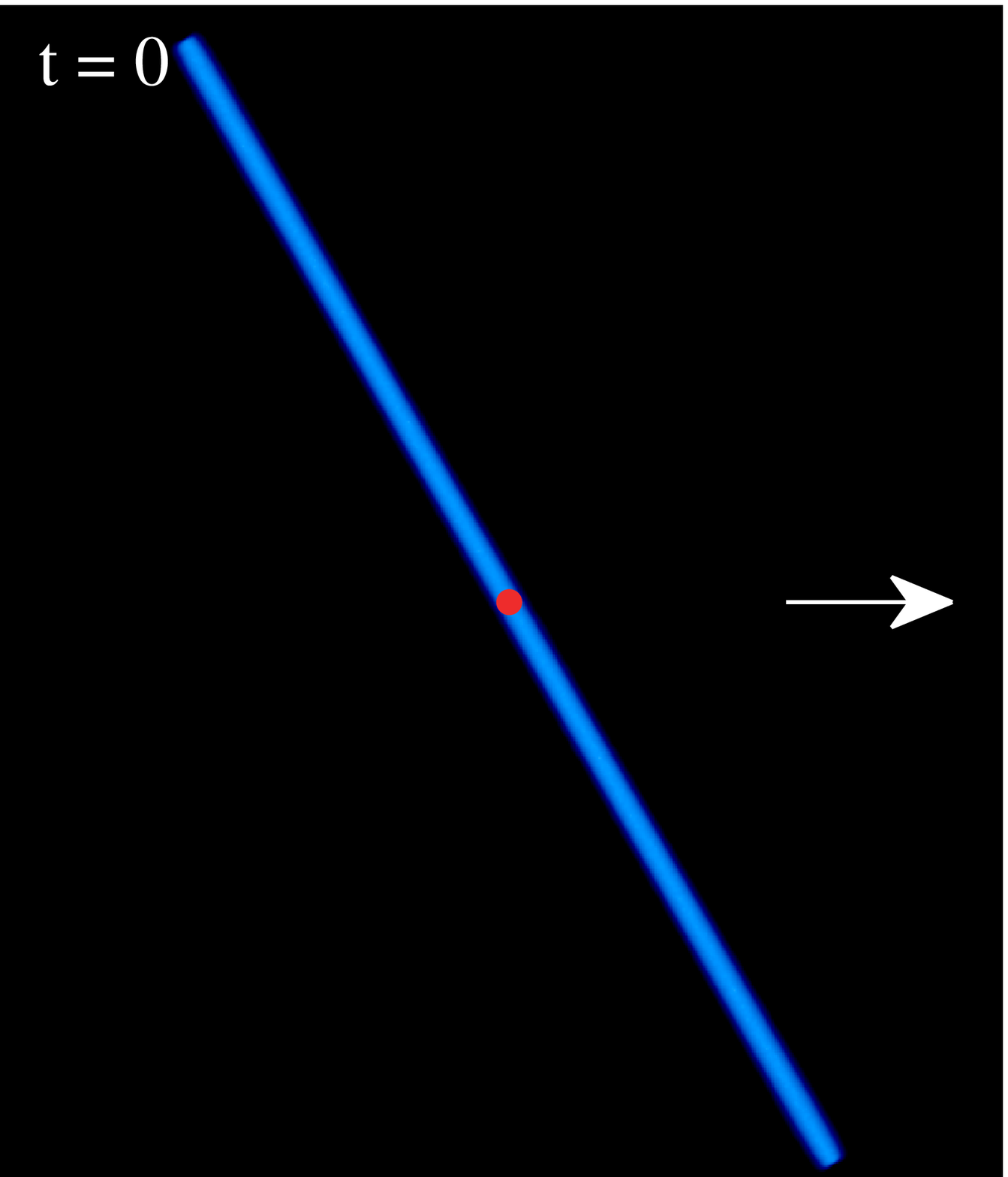}
  \includegraphics[angle=0,width=0.35\textwidth]{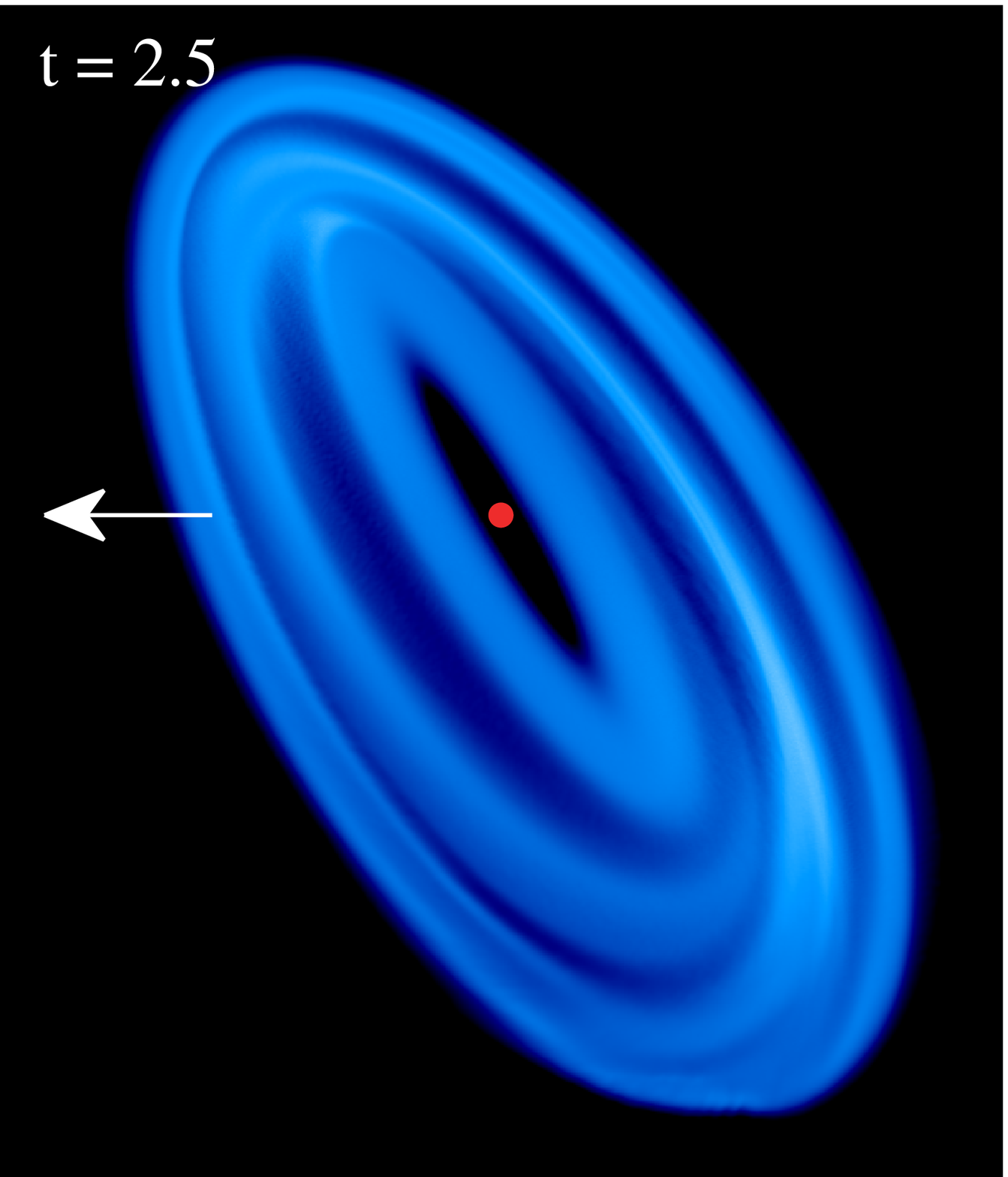}
  \includegraphics[angle=0,width=0.35\textwidth]{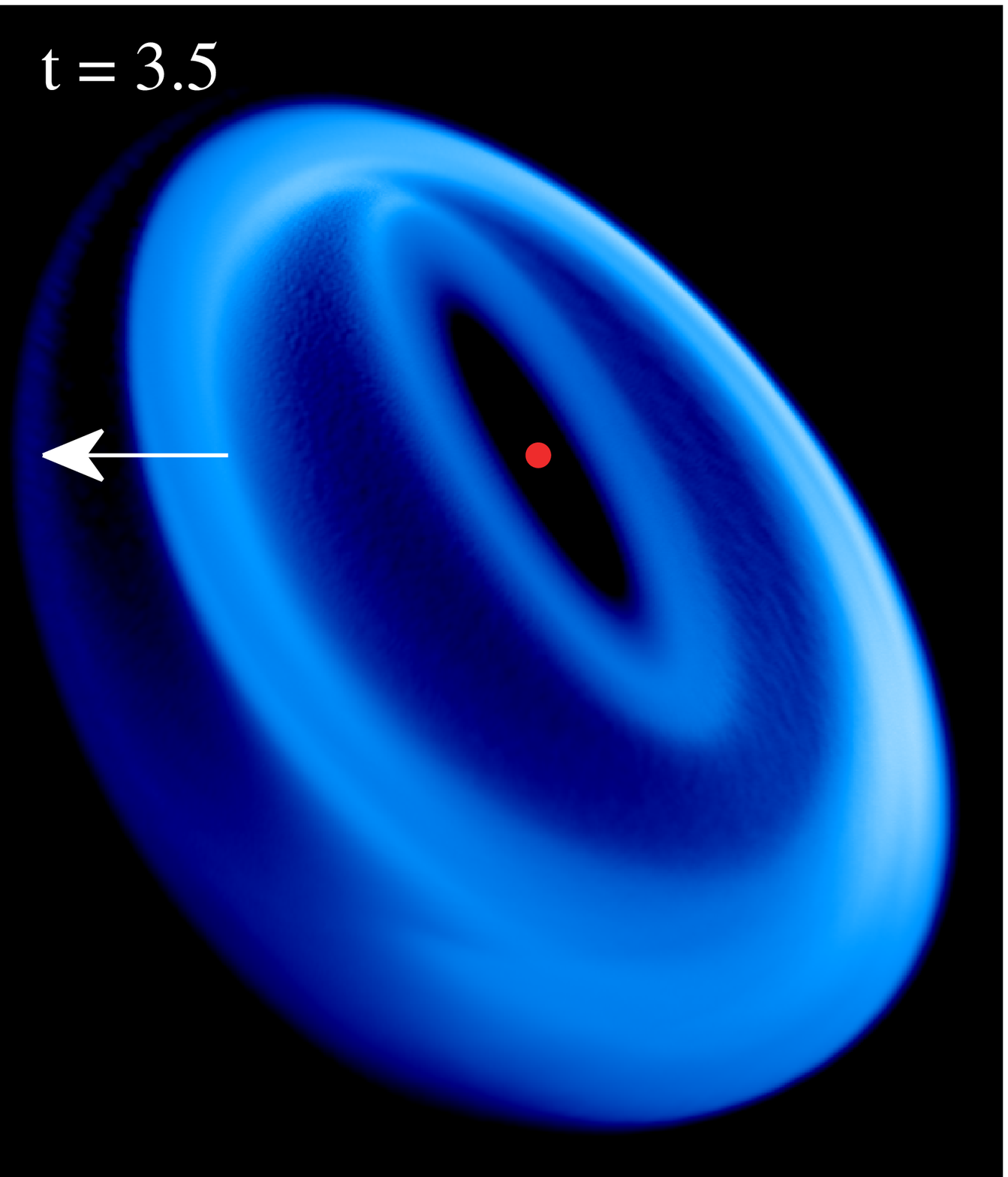}
  \includegraphics[angle=0,width=0.35\textwidth]{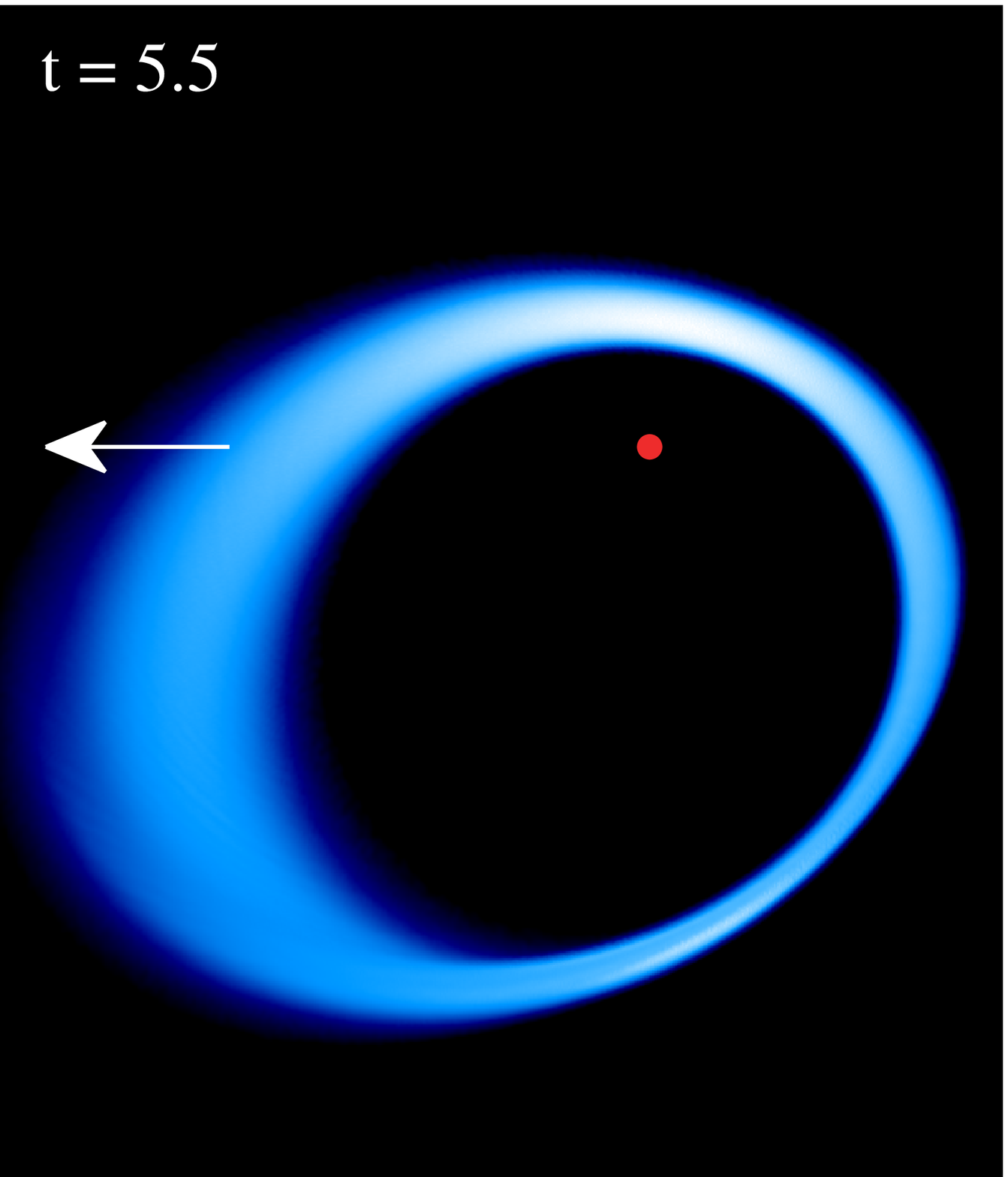}
  \caption{3D surface rendering of the disc which was initially inclined at $60^{\circ}$ to the binary plane with no warp. These snapshots are taken after 0, 2.5, 3.5 and 5.5 binary orbits. The disc is viewed along the binary orbital plane and the arrow points in the direction of the companion in the snapshots. Four stages are shown: (i) the initial inclined disc, (ii) the disc is disrupted by the precession torque, (iii) the disc breaks into two distinct rings, but the outer ring is growing more eccentric and (iv) the outer ring and inner ring merge, causing enhanced dissipation and leaving behind an inclined eccentric disc.}
  \label{tilt60}
\end{figure*}

To check our analytical reasoning above, we perform 3D hydrodynamical numerical simulations using the \textsc{phantom} smoothed particle hydrodynamics code \citep{Price2012a}, as in previous studies of warped discs (e.g. \citealt{LP2010}; \citealt{Nixon2012}) and broken discs (\citealt{Nixonetal2012b}; \citealt{Nixonetal2013}). Disc breaking has also been observed in the circumbinary simulations of \cite{LP1997}, and the unforced warped--disc simulations of \cite{LP2010}.

We follow the method of \cite{Nixonetal2013}, but simulate discs around one component of the binary rather than circumbinary discs. The specific parameters used here can be summarized as follows: The disc is initially planar and extends from $R_{\rm in}=0.1a$ to $R_{\rm out}=0.35a$ with a surface density profile $\Sigma = \Sigma_0(R/R_{\rm in})^{-p}$ and locally isothermal sound speed profile $c_{\rm s} = c_{{\rm s,}0}(R/R_{\rm in})^{-q}$, where we have chosen $p = 3/2$ and $q = 3/4$. This achieves a uniformly resolved disc with the shell-averaged smoothing length per disc scale-height $\left<h\right>/H \approx$ constant \citep{LP2007}. $\Sigma_0$ and $c_{{\rm s,}0}$ are set by the disc mass, $M_{\rm d}=10^{-3}M$ and the disc angular semi-thickness, $H/R = 0.01$ (at $R=R_{\rm in}$) respectively. Initially the disc is composed of 4 million particles, which for this setup gives $\left<h\right>/H \approx 0.5$. The simulations use a disc viscosity with Shakura \& Sunyaev $\alpha \simeq$ 0.1 (which requires artificial viscosity $\alpha_{\rm AV} = 1.9$; cf. \citealt{LP2010}) and a quadratic artificial viscosity $\beta_{\rm AV}=2$. We assume that the binary components, represented by two Newtonian point masses with $M_1=M_2=0.5M$, accrete any gas coming within a distance $0.05a$ of them, and so remove this gas from the computation. The simulations differ only in the initial inclination angle between the disc and the binary orbit. We perform our simulations for $\theta = 10^{\circ}, 30^{\circ}, 45^{\circ}$ and $60^{\circ}$.

\begin{figure}
  \includegraphics[angle=0,width=\columnwidth]{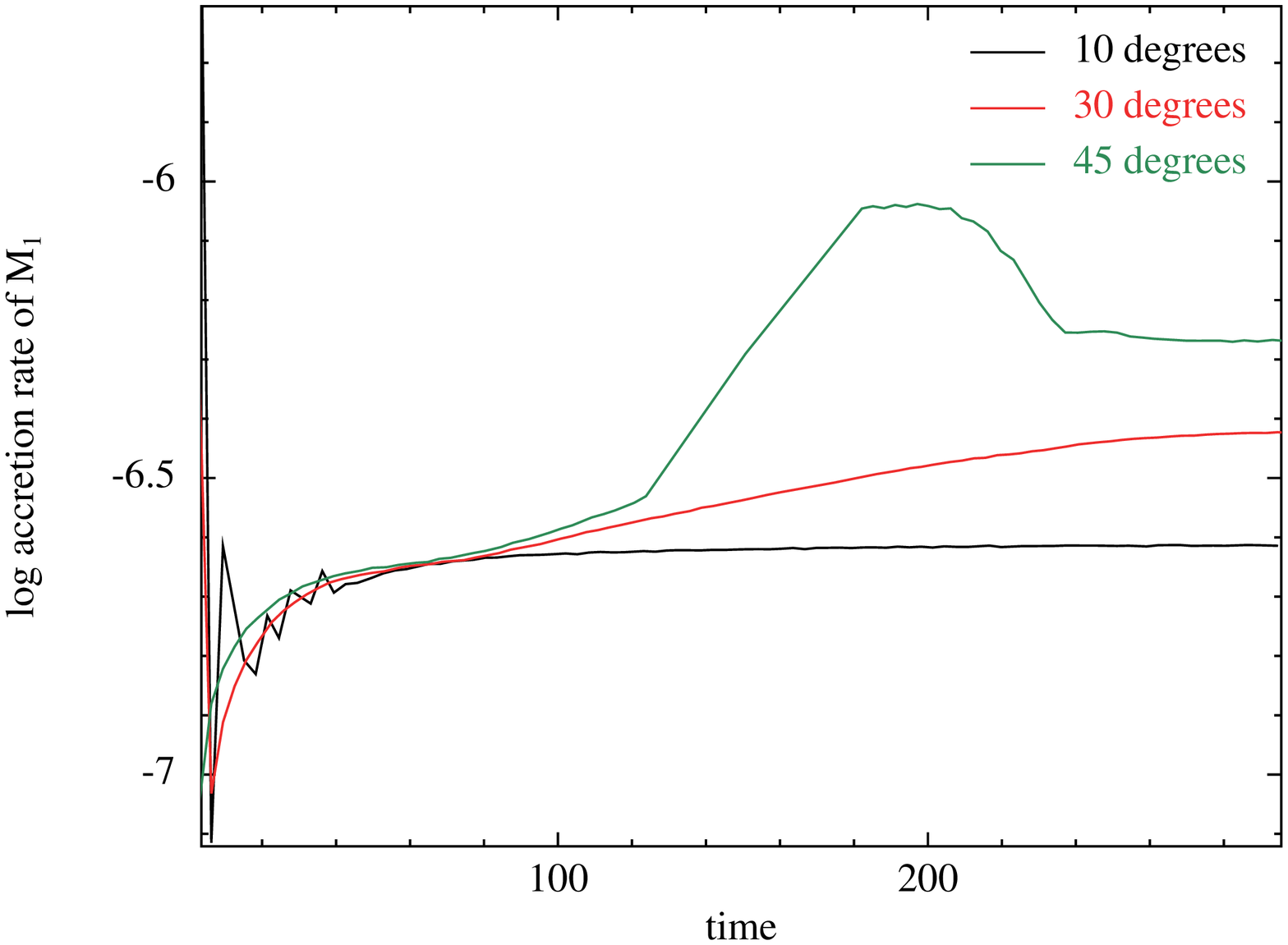}\vspace{0.2in}
  \includegraphics[angle=0,width=\columnwidth]{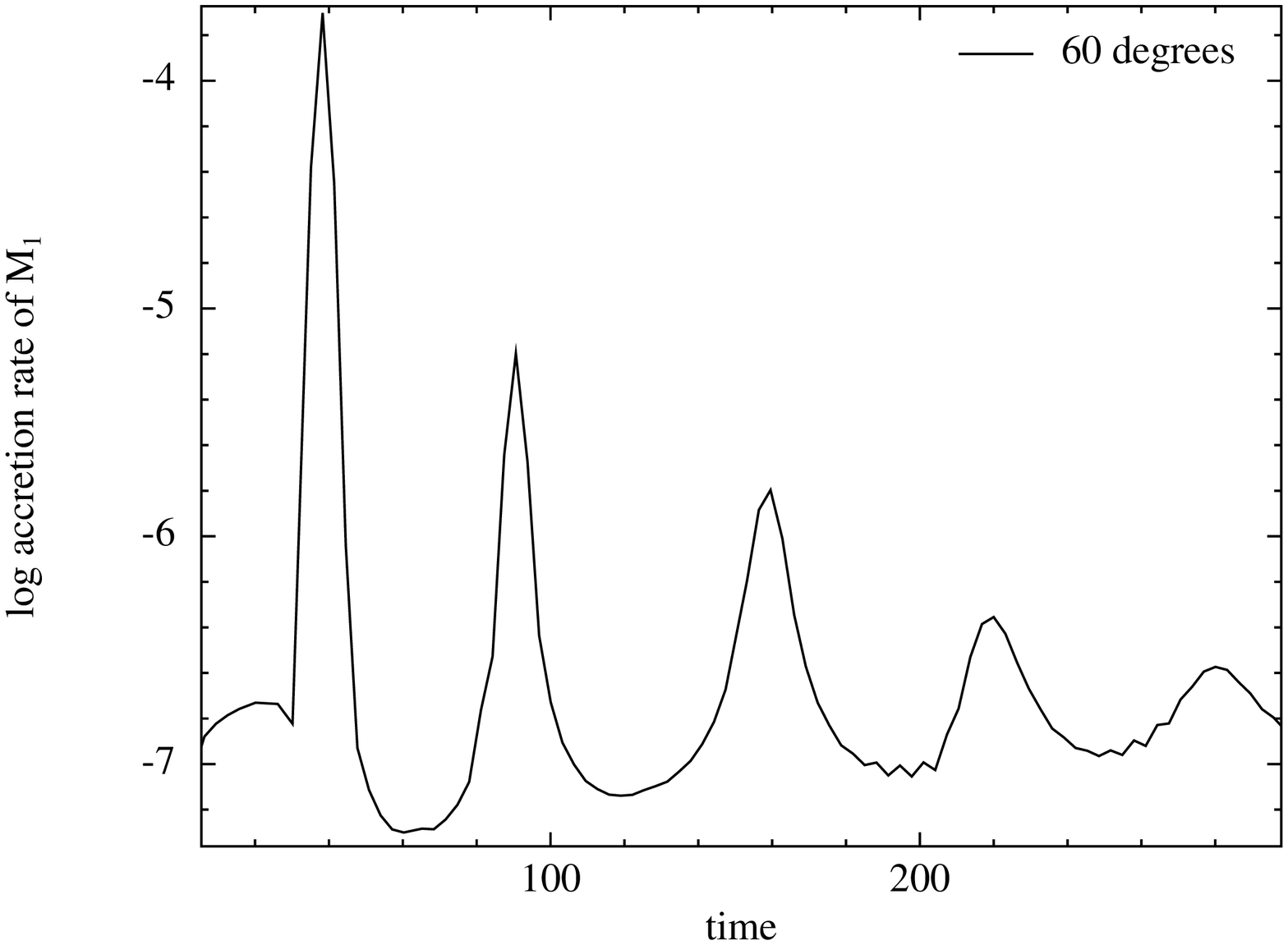}
  \caption{The accretion rates with time for different inclination angles of the disc. The accretion rate is calculated in time bins of half a binary orbit. The accretion rate is in code units of binary mass per binary dynamical time, and the time axis is in units of the binary dynamical time where a full orbit of the binary is $2\pi$.}
  \label{acc}
\end{figure}

Fig.~\ref{tilt10} shows the simulation with an initial inclination of $10^{\circ}$. Here the precession torque caused by the companion is weak, so the disc evolves with a mild warp. We know from Eq.~\ref{gp} that the strength of the precession torque is higher when $\theta=30^{\circ}$, and it has its maximum value when $\theta=45^{\circ}$. This agrees with  Eq.~\ref{rbreak} which shows that disc breaking is more likely when $\sin 2\theta$ is high. We find strong disc breaking in our simulations with initial inclinations of $30^{\circ}$ and $45^{\circ}$. Fig.~\ref{tilt30} shows a simulation with an initial inclination of $30^{\circ}$. Here the disc becomes significantly warped after a few orbits. Then the outer disc breaks off to form a distinct outer ring. Similarly, the disc with an initial inclination of $45^{\circ}$ is disrupted by the strong precession torque and initially breaks into two distinct planes. Then a third ring is broken off, but quickly interacts with the outer ring. The two outermost rings merge after another $\sim$10 binary orbits, as shown in Fig.~\ref{tilt45}. The inner ring develops a strong warp, aligning somewhat towards the binary plane, while the outer ring remains highly inclined and precessing. The $\theta =60^{\circ}$ simulation evolves quite differently from those with smaller $\theta$. This is shown in Fig.~\ref{tilt60}. The simulated disc appears as if it is about to tear after a few binary orbits, but the outer regions of the disc become quite eccentric. The inner and outer disc interact strongly at pericentre, causing enhanced dissipation (cf. Fig.~\ref{acc}) and merging the two rings into a single eccentric disc. The remaining eccentric disc persists over the duration of the simulation, which is approximately 50 binary orbits.

The simplified criterion (\ref{rbreak}) derived in Section~\ref{tearing} predicts the breaking found in the simulations with inclination angle $\ge 30^\circ$. However, it suggests that the disc with inclination angle $10^\circ$ should also break with $R_{\rm break} \gtrsim 0.23$, within the disc outer radius ($R_{\rm out}=0.35$). Instead of breaking the disc, we find that the $10^\circ$ simulation aligns to the binary plane after $\sim 20$ binary orbits. This alignment suggests that the vertical viscous torque (\ref{gnu2}) is dynamically important on short timescales. In this case, the simplifications made between (\ref{rbreakt}) and (\ref{rbreak}) are not relevant. We also note that the outer disc radius in the $10^\circ$ simulation shrinks slightly to $R_{\rm out} \approx 0.3a$ due to disc--binary resonances \citep{AL1994}.

The importance of the vertical viscous torque for the $10^\circ$ simulation can easily be shown by estimating the $\alpha_2$ term in Eq. \ref{rbreakt}. Analysing this simulation we find the warp amplitude grows to $|\psi| \approx 0.1$ which gives $\alpha_2 \approx 52$ \citep{Ogilvie1999}. From these values we find $R_{\rm break} \gtrsim 0.41a$ by considering the vertical viscosity (see eq. \ref{rbreakt}). The breaking radius predicted by the vertical viscous torque exceeds the disc outer radius and this is in agreement with the simulations. Therefore we can conclude that the simplified criterion (\ref{rbreak}) is not relevant for this case. However, for the $30^\circ$ simulation we find that the warp amplitude grows to $|\psi| \approx 1.5$ which gives $\alpha_2 \approx 0.4$ \citep{Ogilvie1999}. From these values we find that the disc should break with $R_{\rm break} \gtrsim 0.2a$, within the disc outer radius, again in agreement with the simulation. Further we find that the $30^\circ$ simulation breaks at a minimum radius of $0.205a$, whereas the $45^\circ$ simulation breaks at a minimum radius of $0.195a$. From our estimate in Section~\ref{tearing} we expect $R_{\rm break}$ to differ between these two simulations by a factor of $[\sin 60 / \sin 90]^{-1/3} = 1.05$, which is remarkably similar to the difference found in the simulations. These numbers give the smallest radius at which the disc was deemed to have broken, occurring at t = 19 in the $30^\circ$ simulation and t = 16 in the $45^\circ$ simulation.

The accretion rate through tearing discs is generally significantly enhanced. The top panel of Fig.~\ref{acc} shows the accretion rates for the simulations with $\theta=10^{\circ}$, $30^{\circ}$ and $45^{\circ}$. The accretion rate is higher for the broken discs ($\theta=30^{\circ}, 45^{\circ}$) than for the warped disc ($\theta=10^{\circ}$). The disc with $\theta=60^{\circ}$ produces highly variable accretion, as shown in the lower panel of Fig.~\ref{acc}, varying by approximately three orders of magnitude. The high accretion rate for this simulation results from the enhanced dissipation between the inner disc and the eccentric outer regions. The remaining eccentric disc shows a nodding motion which produces the peaks in accretion rate seen in the bottom panel of Fig.~\ref{acc}.

The nodding motion (oscillations in the disc tilt) are accompanied by oscillations in the disc eccentricity. We attribute this to the Kozai--Lidov mechanism recently discovered by \cite{Martinetal2014} to also act in fluid discs. The Kozai--Lidov mechanism induces antiphase oscillations in the orbital inclination and eccentricity of particles highly inclined to a binary companion \citep{Kozai1962,Lidov1962}. \cite{Martinetal2014} are the first to show that this process also occurs in a fluid disc. In our $60^\circ$ simulation the disc quickly becomes a narrow ring (through a strong interaction induced by tearing, which drives particles outside the ring into the accretion radius of the primary). This narrow ring then goes through strong Kozai--Lidov cycles. When the disc has its peak eccentricity the pericentre of the ring approaches the accretion radius and creates the strong accretion rate shown in Fig.~\ref{acc}. \cite{Martinetal2014} were the first to show that this behaviour occurs more generally for extended discs, and is therefore probably important in many astrophysical scenarios.

The increase in accretion rate for the broken discs here is only a factor of a few, rather than the orders of magnitude found in \cite{Nixonetal2013}. The main reason for this is the weaker precession here (cf. eq.~\ref{omprec} compared with \citealt{Nixonetal2013}, eq.~7). In the present case the inner disc evolves significantly faster, reaching a degree of alignment before tearing happens (see Fig.~\ref{tilt45}). This makes the internal disc inclination angle smaller than $ 2\theta$. Another important difference is that resonances hold out a circumbinary disc, but in our case resonances do not slow accretion. We note that for smaller mass ratios ($M_0 < M_1$) the resonances driving superhump behaviour \citep[][]{Whitehurst1988} appear in the disc, and we will investigate this in a further paper. We note finally that significantly higher accretion rates might well appear in the disc for other parameters, e.g. if the viscosity driving alignment was smaller.

\subsection{Waves}
\label{waves}

\begin{figure*}
  \includegraphics[angle=0,width=0.33\textwidth]{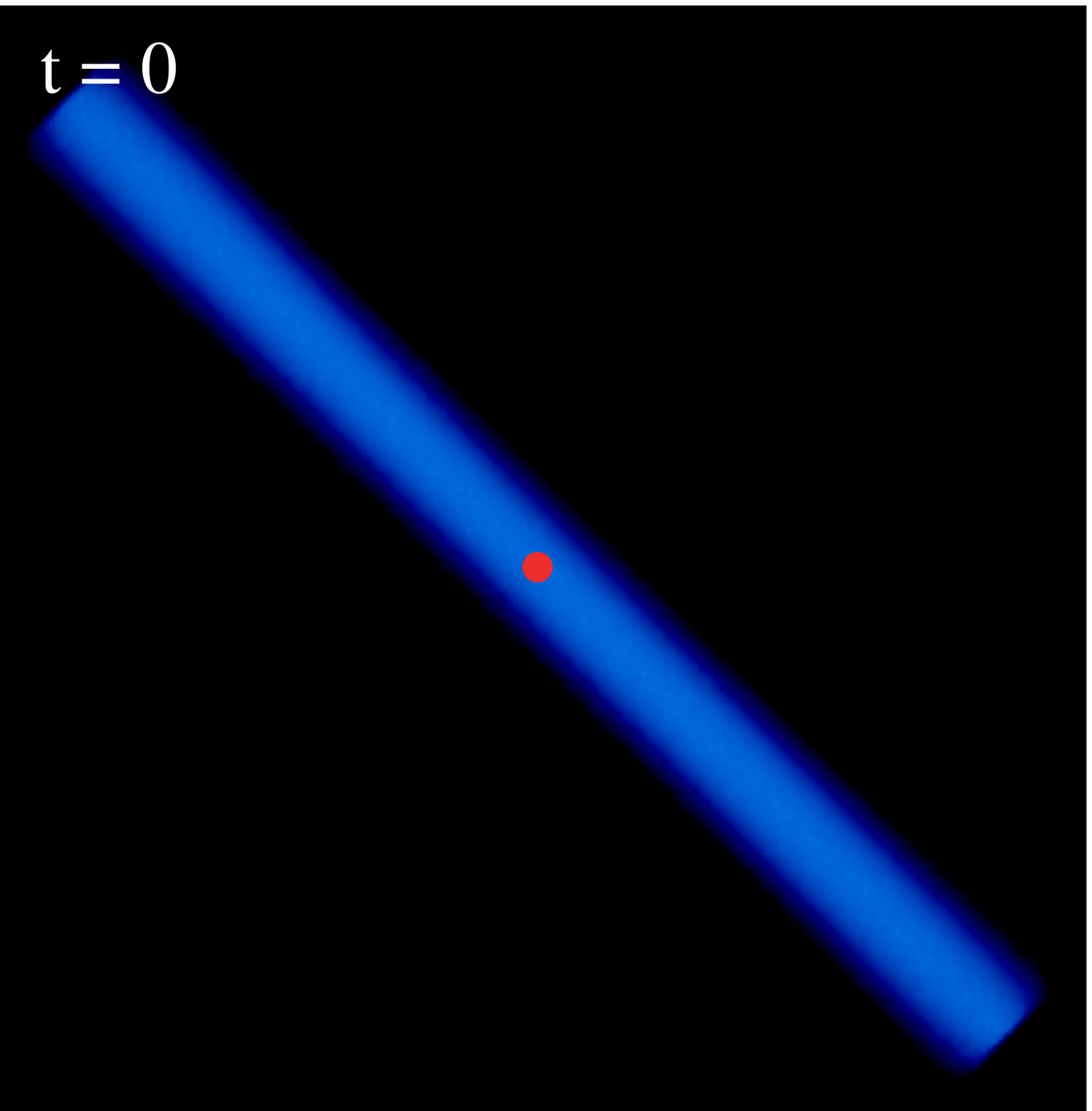}
   \includegraphics[angle=0,width=0.33\textwidth]{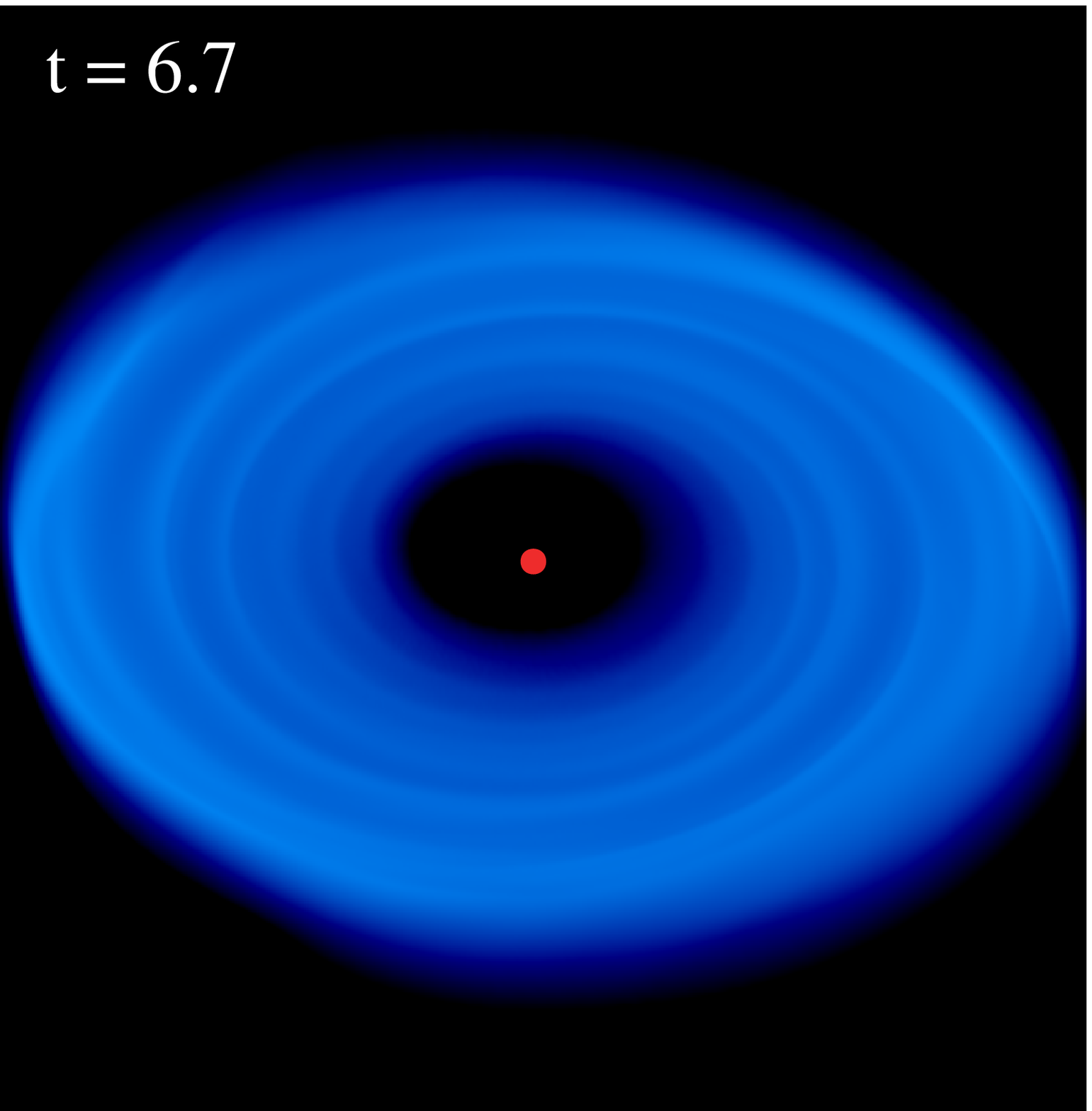}
   \includegraphics[angle=0,width=0.33\textwidth]{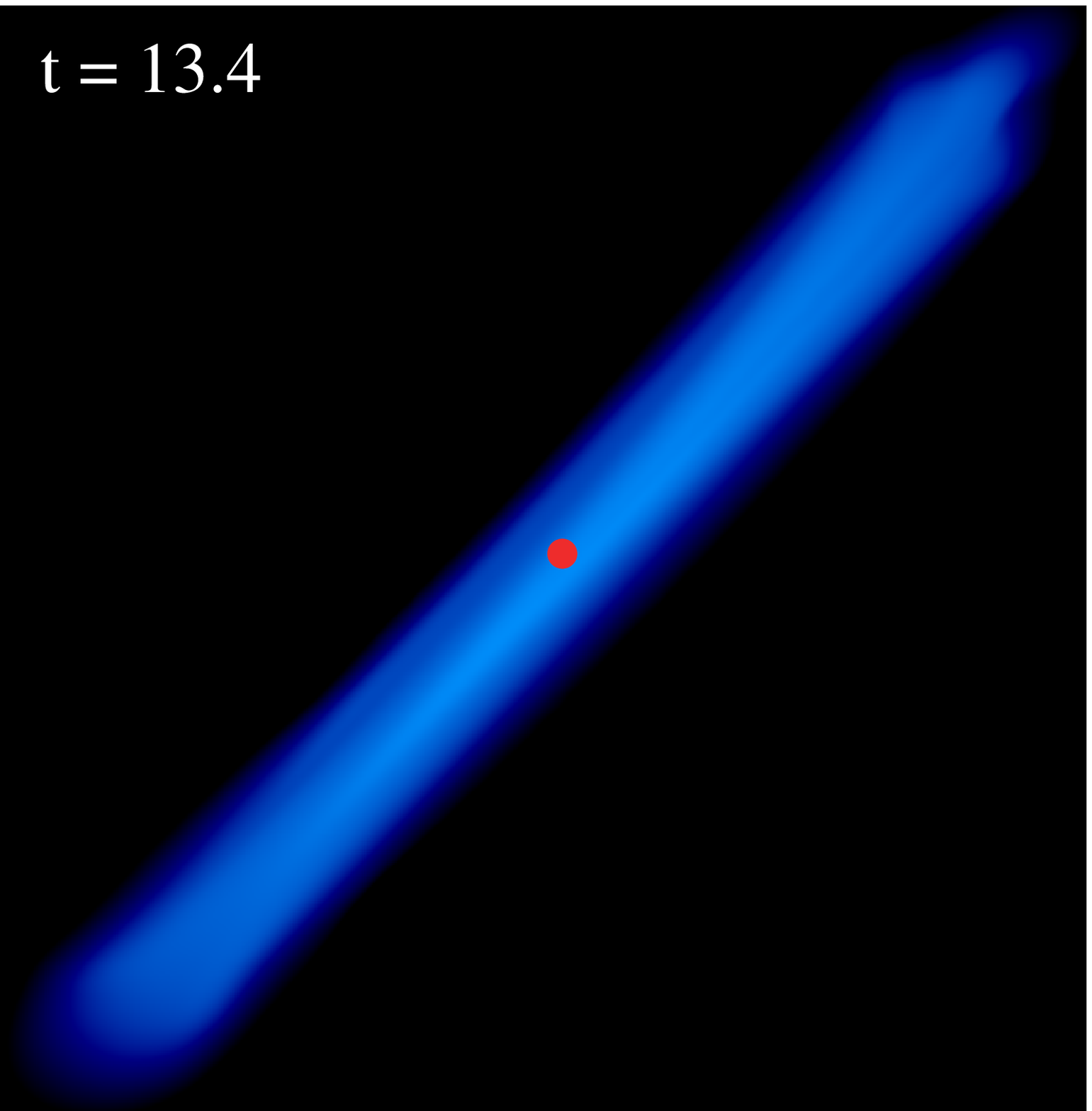}
    \caption{3D surface rendering of a thick ($H/R = 0.05$) disc in the wave--like warp propagation regime ($H/R > \alpha = 0.01$). The disc was initially inclined at $45^{\circ}$ to the binary plane and becomes slightly eccentric over the duration of the simulation. The left hand panel shows the initial conditions and then the middle and right hand panels show the disc after a quarter and a half of a precession period respectively, i.e. a precession of $90^\circ$ and $180^\circ$. The disc is viewed along the binary orbital plane.}
  \label{rigid}
\end{figure*}

In this Section we report a single simulation performed with a thicker disc, $H/R=0.05$ and lower dissipation, $\alpha=0.01$, but otherwise identical parameters to the $45^\circ$ simulation above. For these parameters $\left<h\right>/H = 0.17$ and therefore $\alpha_{\rm AV} = 0.56$. For this simulation, the differential precession induced by the binary is expected to be communicated through pressure waves, propagating at a velocity $v_{\rm w} \approx c_{\rm s}/2$, induced by a warped disc \citep{PL1995,LO2000}. For the parameters of the simulation, the wave travel time across the disc is $t_{\rm w}\approx R_{\rm out}/c_{\rm s}(R_{\rm out}) = 1.3$ binary orbits. However, the fastest precession time induced in the disc is $t_{\rm prec}(R_{\rm out}) = 2\pi/[\Omega_p(R_{\rm out})] = 6.4$ binary orbits, and so we expect the disc to be able to communicate the precession efficiently throughout the disc by pressure waves. This leads to global precession of the disc as seen by previous investigations \citep{Larwoodetal1996}. Fig.~\ref{rigid} shows half a precession period taking $\approx 13$ orbits, so the global precession period observed in the simulation is approximately 25 binary orbits. The predicted global precession time can be calculated by dividing the integral of the angular momentum by the integral of the torque across the disc. For the parameters chosen here, this gives 34 binary orbits, which is consistent with the observed period as the calculation is highly sensitive to the inner and outer disc radii and the surface density profile for our parameter choices and these can change somewhat during the simulation due to viscous spreading of the disc and accretion.

The global precession observed in this simulation reiterates our comment above that the criterion derived in Section~\ref{tearing} should not be applied to scenarios for which the equations are invalid. This simulation is in the wavelike warp propagation regime with $H/R > \alpha$. Therefore the internal disc communication is dominated by waves rather than viscosity. Therefore it is no surprise that this simulation does not agree with (\ref{rbreak}) as this was derived {\it assuming} that warps propagate diffusively. If we instead consider the internal disc communication due to waves, it is clear that the disc should not tear as the wave communication across the whole disc occurs faster than any local precession time in the disc. This is confirmed by our simulation shown in Fig.~\ref{rigid}.

\section{Discussion}
\label{discussion}
We have simulated misaligned accretion discs in a binary system to explore the process of disc tearing, where the precession torque induced in the disc can overwhelm its internal viscous communication. Here a misaligned binary companion gravitationally drives the precession torque. We have shown that sufficiently thin and sufficiently inclined discs can break so that their outer rings precess effectively independently. This process can enhance the dissipation in the disc and promote stronger accretion on to the central object.

In our analytical estimates in Section~\ref{tearing} we compared the precession torque to the internal viscous torques arising from both azimuthal and vertical shear. As it is not straightforward to calculate the vertical torque, which is strongly dependent on warp amplitude, we simplify the full criteria (\ref{rbreakt}) to (\ref{rbreak}). This simplification is relevant for moderate to large values of the disc viscosity parameter $\alpha$ and large inclination angles. We note that an exact calculation of a tearing criterion would require knowledge of $\partial \mathbi{l}/\partial R$ as a function of both position and time, so that the viscosity coefficients and hence the viscous torques can be calculated and compared to the precession torque. For these reasons one has to perform three dimensional hydrodynamic simulations of this process to find out exactly what the disc does. The simplest approach to a criterion for disc tearing is given by (\ref{rbreak}), which has the advantage of being readily calculable, but is not applicable to all regions of parameter space. We shall explore this further with a focussed investigation in a future publication.

We note that the criteria we have derived should not be used when the disc viscosity is smaller than the disc angular semithickness ($\alpha < H/R$) as this allows the efficient propagation of waves \citep{PP1983} and this distinct internal disc communication is not included in our analysis in Section~\ref{tearing}. We have performed one simulation of such a disc with $H/R=0.05$ and $\alpha=0.01$ (Section~\ref{waves}). If we were to naively apply the diffusive tearing criterion (\ref{rbreak}) to this pressure dominated simulation we would expect the disc to tear similarly to Figs~\ref{tilt45} \& \ref{tilt60}. However, we instead find that the disc precesses as a solid body. This happens because the differential precession induced by the binary is communicated across the whole disc by pressure waves, which propagate at a velocity $v_{\rm w} \approx c_{\rm s}/2$ \citep{PL1995,LO2000}. This leads to global precession of the disc, as seen in previous investigations \citep{Larwoodetal1996}. We note that if we consider a criterion which takes account of wave communication in the disc \citep{Nixonetal2013}, then we instead predict that the disc should not break, consistent with this simulation.

\section{Conclusions}
\label{conclusions}
We have shown that tilted discs inside a binary are susceptible to tearing from the outside in, because of the gravitational torque from the companion star. If the disc inclination is small the disc warps (Fig.~\ref{tilt10}), and if there is nothing to maintain the tilt, the disc eventually aligns with the binary plane. For larger inclinations the disc can be torn, the outer ring breaking off and precessing effectively independently (Figs.~\ref{tilt30} \& \ref{tilt45}). For some inclinations the eccentricity of the outer disc grows (Fig.~\ref{tilt60}).

The behaviour we have discussed in this paper is relevant to a variety of astrophysical systems, for example X--ray binaries, where the disc plane may be tilted by radiation warping (e.g. \citealt{WP1999}; \citealt{OD2001}), supermassive black hole binaries, where accretion of misaligned gas can create effectively random inclinations \citep[e.g.][]{Nixonetal2013} and protostellar binaries, where a disc may be misaligned by a variety of effects such as binary capture/exchange, accretion after binary formation \citep{Bateetal2010} and stellar flybys \citep{NP2010}.

\section*{Acknowledgments}
We thank Phil Armitage for useful discussions and the referee for useful comments. SD gratefully acknowledges the warm hospitality of the Theoretical Astrophysics Group at University of Leicester during her visits. SD is supported for this work by The Scientific and Technological Research Council of Turkey (T\"{U}B\.{I}TAK) through the Postdoctoral Research Fellowship Programme (2219). CJN was supported for this work by NASA through the Einstein Fellowship Program, grant PF2-130098. DJP is supported by a Future Fellowship (FT130100034) from the Australian Research Council. Research in theoretical astrophysics at Leicester is supported by an STFC Consolidated Grant. We used {\sc splash} \citep{Price2007} for the visualization. The calculations for this paper were performed on the Complexity node of the DiRAC2 HPC Facility which is jointly funded by STFC, the department of Business Innovation and Skills and the University of Leicester.

\bibliographystyle{mn2e}
\bibliography{nixon}

\begin{thebibliography}{}

\bibitem[\protect\citeauthoryear{{Artymowicz} \& {Lubow}}{{Artymowicz} \&
  {Lubow}}{1994}]{AL1994}
{Artymowicz} P.,  {Lubow} S.~H.,  1994, \apj, 421, 651

\bibitem[\protect\citeauthoryear{{Balbus} \& {Hawley}}{{Balbus} \&
  {Hawley}}{1991}]{BH1991}
{Balbus} S.~A.,  {Hawley} J.~F.,  1991, \apj, 376, 214

\bibitem[\protect\citeauthoryear{{Bardeen} \& {Petterson}}{{Bardeen} \&
  {Petterson}}{1975}]{BP1975}
{Bardeen} J.~M.,  {Petterson} J.~A.,  1975, \apjl, 195, L65+

\bibitem[\protect\citeauthoryear{{Bate}, {Bonnell}, {Clarke}, {Lubow},
  {Ogilvie}, {Pringle} \& {Tout}}{{Bate} et~al.}{2000}]{Bateetal2000}
{Bate} M.~R.,  {Bonnell} I.~A.,  {Clarke} C.~J.,  {Lubow} S.~H.,  {Ogilvie}
  G.~I.,  {Pringle} J.~E.,    {Tout} C.~A.,  2000, \mnras, 317, 773

\bibitem[\protect\citeauthoryear{{Bate}, {Lodato} \& {Pringle}}{{Bate}
  et~al.}{2010}]{Bateetal2010}
{Bate} M.~R.,  {Lodato} G.,    {Pringle} J.~E.,  2010, \mnras, 401, 1505

\bibitem[\protect\citeauthoryear{{Frank}, {King} \& {Raine}}{{Frank}
  et~al.}{2002}]{Franketal2002}
{Frank} J.,  {King} A.,    {Raine} D.~J.,  2002, {Accretion Power in
  Astrophysics: Third Edition}

\bibitem[\protect\citeauthoryear{{Katz}}{{Katz}}{1973}]{Katz1973}
{Katz} J.~I.,  1973, Nature Physical Science, 246, 87

\bibitem[\protect\citeauthoryear{{King}, {Livio}, {Lubow} \& {Pringle}}{{King}
  et~al.}{2013}]{Kingetal2013}
{King} A.~R.,  {Livio} M.,  {Lubow} S.~H.,    {Pringle} J.~E.,  2013, \mnras,
  431, 2655

\bibitem[\protect\citeauthoryear{{King} \& {Pringle}}{{King} \&
  {Pringle}}{2006}]{KP2006}
{King} A.~R.,  {Pringle} J.~E.,  2006, \mnras, 373, L90

\bibitem[\protect\citeauthoryear{{King} \& {Pringle}}{{King} \&
  {Pringle}}{2007}]{KP2007}
{King} A.~R.,  {Pringle} J.~E.,  2007, \mnras, 377, L25

\bibitem[\protect\citeauthoryear{{Kozai}}{{Kozai}}{1962}]{Kozai1962}
{Kozai} Y.,  1962, \aj, 67, 591

\bibitem[\protect\citeauthoryear{{Larwood}, {Nelson}, {Papaloizou} \&
  {Terquem}}{{Larwood} et~al.}{1996}]{Larwoodetal1996}
{Larwood} J.~D.,  {Nelson} R.~P.,  {Papaloizou} J.~C.~B.,    {Terquem} C.,
  1996, \mnras, 282, 597

\bibitem[\protect\citeauthoryear{{Larwood} \& {Papaloizou}}{{Larwood} \&
  {Papaloizou}}{1997}]{LP1997}
{Larwood} J.~D.,  {Papaloizou} J.~C.~B.,  1997, \mnras, 285, 288

\bibitem[\protect\citeauthoryear{{Lidov}}{{Lidov}}{1962}]{Lidov1962}
{Lidov} M.~L.,  1962, \planss, 9, 719

\bibitem[\protect\citeauthoryear{{Lodato} \& {Price}}{{Lodato} \&
  {Price}}{2010}]{LP2010}
{Lodato} G.,  {Price} D.~J.,  2010, \mnras, 405, 1212

\bibitem[\protect\citeauthoryear{{Lodato} \& {Pringle}}{{Lodato} \&
  {Pringle}}{2007}]{LP2007}
{Lodato} G.,  {Pringle} J.~E.,  2007, \mnras, 381, 1287

\bibitem[\protect\citeauthoryear{{Lubow} \& {Ogilvie}}{{Lubow} \&
  {Ogilvie}}{2000}]{LO2000}
{Lubow} S.~H.,  {Ogilvie} G.~I.,  2000, \apj, 538, 326

\bibitem[\protect\citeauthoryear{{Martin}, {Nixon}, {Lubow}, {Armitage},
  {Price}, {Do{\u g}an} \& {King}}{{Martin} et~al.}{2014}]{Martinetal2014}
{Martin} R.~G.,  {Nixon} C.,  {Lubow} S.~H.,  {Armitage} P.~J.,  {Price} D.~J.,
   {Do{\u g}an} S.,    {King} A.,  2014, \apjl, 792, L33

\bibitem[\protect\citeauthoryear{{Nixon}, {King} \& {Price}}{{Nixon}
  et~al.}{2013}]{Nixonetal2013}
{Nixon} C.,  {King} A.,    {Price} D.,  2013, \mnras, 434, 1946

\bibitem[\protect\citeauthoryear{{Nixon}, {King}, {Price} \& {Frank}}{{Nixon}
  et~al.}{2012}]{Nixonetal2012b}
{Nixon} C.,  {King} A.,  {Price} D.,    {Frank} J.,  2012, \apjl, 757, L24

\bibitem[\protect\citeauthoryear{{Nixon} \& {Salvesen}}{{Nixon} \&
  {Salvesen}}{2014}]{NS2014}
{Nixon} C.,  {Salvesen} G.,  2014, \mnras, 437, 3994

\bibitem[\protect\citeauthoryear{{Nixon}}{{Nixon}}{2012}]{Nixon2012}
{Nixon} C.~J.,  2012, \mnras, 423, 2597

\bibitem[\protect\citeauthoryear{{Nixon} \& {King}}{{Nixon} \&
  {King}}{2012}]{NK2012}
{Nixon} C.~J.,  {King} A.~R.,  2012, \mnras, 421, 1201

\bibitem[\protect\citeauthoryear{{Nixon} \& {Pringle}}{{Nixon} \&
  {Pringle}}{2010}]{NP2010}
{Nixon} C.~J.,  {Pringle} J.~E.,  2010, \mnras, 403, 1887

\bibitem[\protect\citeauthoryear{{Ogilvie}}{{Ogilvie}}{1999}]{Ogilvie1999}
{Ogilvie} G.~I.,  1999, \mnras, 304, 557

\bibitem[\protect\citeauthoryear{{Ogilvie}}{{Ogilvie}}{2000}]{Ogilvie2000}
{Ogilvie} G.~I.,  2000, \mnras, 317, 607

\bibitem[\protect\citeauthoryear{{Ogilvie}}{{Ogilvie}}{2003}]{Ogilvie2003}
{Ogilvie} G.~I.,  2003, \mnras, 340, 969

\bibitem[\protect\citeauthoryear{{Ogilvie} \& {Dubus}}{{Ogilvie} \&
  {Dubus}}{2001}]{OD2001}
{Ogilvie} G.~I.,  {Dubus} G.,  2001, \mnras, 320, 485

\bibitem[\protect\citeauthoryear{{Papaloizou} \& {Lin}}{{Papaloizou} \&
  {Lin}}{1995}]{PL1995}
{Papaloizou} J.~C.~B.,  {Lin} D.~N.~C.,  1995, \apj, 438, 841

\bibitem[\protect\citeauthoryear{{Papaloizou} \& {Pringle}}{{Papaloizou} \&
  {Pringle}}{1983}]{PP1983}
{Papaloizou} J.~C.~B.,  {Pringle} J.~E.,  1983, \mnras, 202, 1181

\bibitem[\protect\citeauthoryear{{Petterson}}{{Petterson}}{1977a}]{Petterson1977b}
{Petterson} J.~A.,  1977a, \apj, 218, 783

\bibitem[\protect\citeauthoryear{{Petterson}}{{Petterson}}{1977b}]{Petterson1977a}
{Petterson} J.~A.,  1977b, \apj, 216, 827

\bibitem[\protect\citeauthoryear{{Price}}{{Price}}{2007}]{Price2007}
{Price} D.~J.,  2007, PASA, 24, 159

\bibitem[\protect\citeauthoryear{{Price}}{{Price}}{2012}]{Price2012a}
{Price} D.~J.,  2012, Journal of Computational Physics, 231, 759

\bibitem[\protect\citeauthoryear{{Pringle}}{{Pringle}}{1981}]{Pringle1981}
{Pringle} J.~E.,  1981, \araa, 19, 137

\bibitem[\protect\citeauthoryear{{Pringle}}{{Pringle}}{1992}]{Pringle1992}
{Pringle} J.~E.,  1992, \mnras, 258, 811

\bibitem[\protect\citeauthoryear{{Pringle}}{{Pringle}}{1996}]{Pringle1996}
{Pringle} J.~E.,  1996, \mnras, 281, 357

\bibitem[\protect\citeauthoryear{{Pringle}}{{Pringle}}{1997}]{Pringle1997}
{Pringle} J.~E.,  1997, \mnras, 292, 136

\bibitem[\protect\citeauthoryear{{Pringle} \& {Rees}}{{Pringle} \&
  {Rees}}{1972}]{PR1972}
{Pringle} J.~E.,  {Rees} M.~J.,  1972, \aap, 21, 1

\bibitem[\protect\citeauthoryear{{Shakura} \& {Sunyaev}}{{Shakura} \&
  {Sunyaev}}{1973}]{SS1973}
{Shakura} N.~I.,  {Sunyaev} R.~A.,  1973, \aap, 24, 337

\bibitem[\protect\citeauthoryear{{Torkelsson}, {Ogilvie}, {Brandenburg},
  {Pringle}, {Nordlund} \& {Stein}}{{Torkelsson}
  et~al.}{2000}]{Torkelssonetal2000}
{Torkelsson} U.,  {Ogilvie} G.~I.,  {Brandenburg} A.,  {Pringle} J.~E.,
  {Nordlund} {\AA}.,    {Stein} R.~F.,  2000, \mnras, 318, 47

\bibitem[\protect\citeauthoryear{{Whitehurst}}{{Whitehurst}}{1988}]{Whitehurst1988}
{Whitehurst} R.,  1988, \mnras, 232, 35

\bibitem[\protect\citeauthoryear{{Wijers} \& {Pringle}}{{Wijers} \&
  {Pringle}}{1999}]{WP1999}
{Wijers} R.~A.~M.~J.,  {Pringle} J.~E.,  1999, \mnras, 308, 207

\end{thebibliography}

\label{lastpage}
\end{document}